\documentclass[12pt,preprint]{aastex}

\usepackage{natbib}
\usepackage{amsmath,amsfonts,amssymb}
\usepackage{graphicx}
\usepackage{wasysym}
\usepackage{float}
\usepackage{url}
\newcommand{\be}{\begin{equation}}
\newcommand{\ee}{\end{equation}}
\newcommand \beq {\begin{equation}}
\newcommand \eeq {\end{equation}}

\newcommand{\ik}{{\it Kepler~}}
\newcommand{\ikt}{{\it Kepler}}

\slugcomment{Invited Review for \textit{Treatise on Geophysics}, 2nd Edition}
\shorttitle{Exoplanetary Geophysics}

\title{Exoplanetary Geophysics -- An Emerging Discipline}

\author{Gregory Laughlin}
\affil{UCO/Lick Observatory, University of California, Santa Cruz, Santa Cruz, CA 95064, USA}

\author{Jack J. Lissauer}
\affil{NASA Ames Research Center, Planetary Systems Branch, Moffett Field, CA 94035, USA}

\begin{document}
\maketitle

\section{Abstract}

Thousands of extrasolar planets have been discovered, and it is clear that the galactic planetary census draws on a diversity greatly exceeding that exhibited by the solar system's planets. We review significant landmarks in the chronology of extrasolar planet detection, and we give an overview of the varied observational techniques that are brought to bear. We then discuss the properties of the planetary distribution that is currently known, using the mass-period diagram as a guide to delineating hot Jupiters, eccentric giant planets, and a third, highly populous, category that we term ``ungiants'', planets having masses $M<30\,M_{\oplus}$ and orbital periods $P<100\,{\rm d}$. We then move to a discussion of the bulk compositions of the extrasolar planets, with particular attention given to the distribution of planetary densities. We discuss the long-standing problem of radius anomalies among giant planets, as well as issues posed by the unexpectedly large range in sizes observed for planets with mass somewhat greater than Earth's. We discuss the use of transit observations to probe the atmospheres of extrasolar planets; various measurements taken during primary transit, secondary eclipse, and through the full orbital period, can give clues to the atmospheric compositions, structures and meteorologies. The extrasolar planet catalog, along with the details of our solar system and observations of star-forming regions and protoplanetary disks, provide a backdrop for a discussion of planet formation in which we review the elements of the favored pictures for how the terrestrial and giant planets were assembled. We conclude by listing several research questions that are relevant to the next ten years and beyond.

\newpage
\section{Introduction}

Until nearly the close of the twentieth century, the search for planets beyond the Solar System was a fraught and generally less-than-fully respectable endeavor. For generations, the scientific community was subjected to spurious announcements of discovery of planets orbiting nearby stars. Famous examples include erroneous claims, starting in 1855 of a planetary mass object in the 70 Ophiuchi system \citep[e.g.][]{1855MNRAS..15..228J, 1896AJ.....16...17S}, and that  one, or possibly two, Jupiter-like planets were harbored by Barnard's Star, the red dwarf that is our Sun's nearest stellar neighbor aside from the $\alpha$ Centauri triple-star system \citep{1963AJ.....68..515V, 1969AJ.....74..757V}.

If we adopt the frequently-used definition that requires the planetary mass, $M_{\rm P}$, to lie below the 13 M$_{\rm J}$ deuterium-burning limit, where M$_{\rm J}$ denotes the mass of Jupiter, then it might be appropriate to credit the first bona-fide detection of an extrasolar planet to \citet{1989Natur.339...38L}, who used Doppler radial velocity measurements to sense an object with a minimum mass of 11 Jupiter masses on an eccentric 83.9-day orbit around HD 114762, a relatively nearby, solar-type star. With hindsight, this object belongs to the class of giant planets that tend to be more massive than Jupiter, which often have substantial orbital eccentricities, and which are found in orbit around approximately 5\% of local solar-type dwarfs \citep{2006ApJ...646..505B}. At the time of HD 114762 b's discovery, however, as a consequence of the abysmal track record of extrasolar planet claims, Latham et al.'s announcement  did not generate significant excitement within the wider community.  Indeed, the profound difference between HD 114762 b's mass and orbit from those of our Solar System's planets led Latham et al.~to state ``This leads to the suggestion that the companion is probably a brown dwarf, and may even be a giant planet." Given that HD 114762 b's product of its mass and sine of the inclination angle between its orbital plane, $M_{\rm P} \sin i$, is so near this physical mass boundary and $\sin i < 1$, Latham et al.'s statement can still be considered as a proper assessment of our current knowledge of this object a quarter century later! 

Soon thereafter, astronomers developed a sudden and abiding interest in extrasolar planets. The catalyst was the pulse-timing detection, by \citet{1992Natur.355..145W}, of two terrestrial-mass planets, with $M_{\rm P} \sin i$ = 3.4 M$_{\oplus}$ and 2.8 M$_{\oplus}$, and with Mercury-like periods of 66.5 and 98.2 days, orbiting the millisecond pulsar PSR B1257+12.  Pulsars, which are rapidly rotating, strongly magnetized neutron stars, emit radio waves that appear as periodic pulses to an observer on Earth. Although a pulsar emits an exquisitely periodic signal, the times at which the pulses reach the receiver are not equally spaced if the distance between the pulsar and the telescope varies in a nonlinear fashion, and if periodic variations are present in data that has been properly reduced, they may indicate the presence of companions orbiting the pulsar. In the case of  PSR B1257+12, measurements are so accurate that deviations from keplerian orbits of the two planets mentioned above caused by their mutual interactions have been measured and used to calculate their actual masses, $M_{\rm P}$ = 4.3 $\pm$ 0.2 M$_{\oplus}$ and 3.9 $\pm$ 0.2 M$_{\oplus}$, implying that the system is inclined by $\sim 50^\circ$ from the line of sight \citep{2003ApJ...591L.147K}. A handful of other pulsar planets have been found, including a  lunar-mass third planet orbiting PSR B1257+12, and a long-period $\sim2.5M_{\rm J}$ planet in a circumbinary orbit about a pulsar-white dwarf binary system in the M4 globular cluster \citep{2003Sci...301..193S}.

The pulsar planets were followed in 1995 by Michel Mayor and Didier Queloz's discovery of 51 Pegasi b,\footnote{``Official'' exoplanetary names are a compound of the name of the parent star and a lower-case letter. The first planet detected in orbit around a given star is denoted {\it b}, the second, {\it c}, etc.} a gas giant planet with a minimum mass roughly half that of Jupiter, and a startlingly short 4.2308-day orbital period about a metal-rich, but otherwise ordinary G-type main sequence star. Thereafter, the pace of discovery ballooned, along with our understanding of both the planetary census and of the physical characteristics of alien worlds. The study of exoplanet physical characteristics has largely been made possible by the detection of transiting planets, which partially eclipse the light of their star once per orbit, the first of  which was discovered in 1999 in orbit around HD 209458 \citep{2000ApJ...529L..41H, Charbonneau00}. 

It is no exaggeration to say that a fully independent branch of astronomy has been created from scratch. The rapid pace of progress is reflected in Figure 1, which charts $\log_{10} M_{\rm P} \sin i$ as a function of year of discovery. The lower envelope of the discovery masses in this semi-log plot is well described by a constant slope, indicative of a Moore's-law behavior with a $\tau \sim$ 2 year halving time of the lowest-known extrasolar planetary mass.

\begin{figure}[H]
\centering
\includegraphics[height=14cm,angle=0]{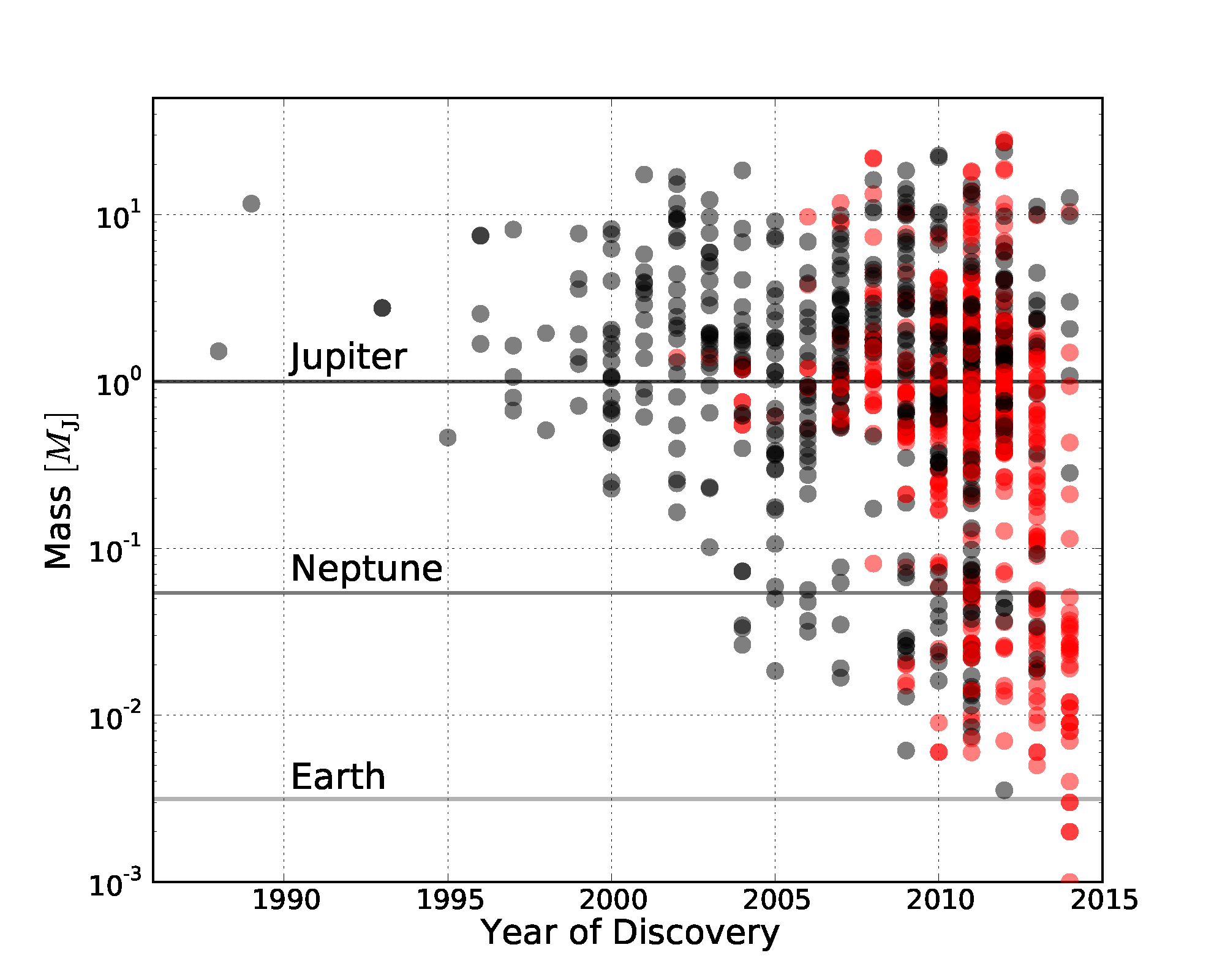}
\caption{Discovery date vs. estimated masses for extrasolar planets orbiting normal stars (not pulsars) over time, with a logarithmic $y$-axis showing estimated masses (or for the cases of Doppler detections, $M \sin i$'s), and year of discovery running along the $x$-axis. The lower envelope of this semi-log plot shows a clear Moore's Law-like progression, and demonstrates that the field has collectively arrived (in the past 1-2 years) at the discovery of planets with $\sim 1$ M$_{\oplus}$. Transiting planets are shown in red, whereas planets detected \textit{only} via Doppler radial velocity are shown in gray. Data are taken from www.exoplanets.org/csv-files/exoplanets.csv \citep{2014PASP..126..827H}.}
\end{figure}

\section{Planetary Detection Methods}

Given the inherent difficulties associated with the detection of extrasolar planets, it is remarkable that a very broad range of observational methods for characterization and detection have been either successfully employed, or are being seriously prepared for future use. As a corollary to the vast interstellar distances, all extrasolar planets are extremely faint; to a distant optical observer, Jupiter's brightness is $L_{\rm J}\sim 1\times10^{-9}$ the solar value in the V band (although it rises to a contrast of $L_{\rm J}/L_{\odot}\sim2\times10^{-4}$ redward of 30~$\mu$m).  Most detection methods, therefore, are {\it indirect}, in that the planets are sensed through their influence on the host star or on the path of light emitted by a distant star. The range of distinct  observation methods are sensitive to different classes of planets, and importantly, they provide us with complementary information about the planets they do find. Without question, a full quiver of techniques is required in order to provide an optimal characterization of the diversity of planets and of planetary systems.

To date, two methods -- photometric transit detection and Doppler radial velocity measurement -- have generated most of the successful detections. The Doppler radial velocity (RV) technique hinges on the ability to monitor tiny time-dependent wavelength displacements among the profusion of atomic lines in a star's spectrum. By measuring the net overall shift, $z=\delta \lambda/\lambda$, using a large number of individual spectral lines, the velocity at which the star is moving towards or away from an observer can be precisely determined. Astronomers then subtract the motion of the telescope relative to the barycenter of the Solar System and other known motions, revealing the radial motion of the target star that results from the tug of its own planets. Transit photometry identifies planet candidates by the periodic drops in observed stellar brightness that are caused by the planet obscuring a portion of the stellar disk once per orbit.  Note that transit photometry can only detect planets whose orbits are viewed nearly edge-on.

Exoplanet detection using either the radial velocity method or the  transit photometry method requires very precise measurements of light from the planet-hosting star.  The techniques used to achieve such high precision are based on differential measurements of observations taken at different times. Absolute accuracies, therefore, need not be particularly high, so the mean velocity (relative to the Solar System) of a star hosting RV-detected planets is generally not known to an accuracy approaching $1~{\rm m \,s^{-1}}$, nor is the spectrophotometric brightness of a star hosting a planet observed via transit photometry known to the,  e.g., part per thousand level needed in relative brightness to detect Neptune-size planets.

As with pulsar timing, radial velocity measurements yield the product of the planet's mass (divided by that of the star, whose mass can usually be estimated fairly accurately from its spectral characteristics) and the sine of the inclination angle between the orbital plane and the plane of the sky, $M_{\rm P}\sin i$. The total half-amplitude, $K$, of the stellar reflex velocity induced by an orbiting planet is the measured quantity, and is related to $M_{\rm P}\sin i$ through
\begin{equation}
K={\left(\frac{2\pi G}{P}\right)}^{1/3}\frac{M_{\rm P} \sin i}{{(M_{\star}+M_{\rm P})}^{2/3}} \frac{1}{\sqrt{1-e^2}}\, ,
\end{equation}
indicating that the radial velocity technique is most sensitive to massive planets and to planets in short-period orbits.
The RV technique allows estimation of the period, $P$, orbital eccentricity, $e$, the mean anomaly, $M=2\pi(T-T_{\rm periastron})/P$, and orientation -- generally expressed as the longitude of periapse, $\omega$, of the stellar orbit. Multi-parameter estimation techniques have reached a high degree of refinement for the problem of determining best-fit orbital elements and uncertainty estimates. For reviews that cover the process of radial velocity modeling, see \citet{2006ApJ...642..505F, 2009PASP..121.1016M, 2010ApJ...722..937D, 2011A&A...535A..55D}.\footnote{We recommend the {\it Systemic Console} software package, available at \url{http://www.stefanom.org/systemic/}, as an intuitive framework for exploring radial velocity and transit timing data.}

Leading research groups are now routinely achieving instrumentally-based velocity errors of $\sigma \sim50 \,{\rm cm}$~s$^{-1}$ (representing a Doppler shift of two parts in $10^9$) on stable, bright, chromospherically inactive stars. For context, Jupiter causes the Sun's velocity to vary with a half-amplitude $K_{\rm J}=12.5\,{\rm m}\,{\rm s^{-1}}$ and a period $P_{\rm J}=11.86$ years; Saturn's effect is the next largest, with an amplitude of 2.8 ${\rm m}\,{\rm s^{-1}}$ and a period, $P_{\rm S}$, of almost 30 years, but Earth induces a mere $K_{\oplus}=9~{\rm cm}\,{\rm s^{-1}}$ fluctuation. To date, the planet with the lowest radial velocity amplitude signal yet announced is Alpha Centauri B b, with $K=51~{\rm cm}\,{\rm s^{-1}}$, $P=3.2357\,{\rm d}$, and $M \sin i=1.1~$ M$_{\oplus}$ \citep{2012Natur.491..207D}. Earth-like planets orbiting sun-like stars at astrobiologically interesting distances (where liquid water could be stable on their surfaces) remain somewhat beyond the capabilities currently achieved. 

Precise radial velocity measurements require the good statistics that come from aggregating individual measurements of a large number of spectral lines. High precision measurements thus cannot be achieved for hotter stars with temperatures $T_{\rm eff}\gtrsim6000$ K (corresponding to spectral types O, B, A, and early F), as these stars have far fewer spectral features than do stars of solar type or cooler. Stellar rotation and intrinsic variability (including star spots) also represent major sources of noise that confound radial velocity measurements. 
The radial velocity surveys, such as the Geneva Planet Search \citep{2009A&A...493..639M}, the California Planet Search \citep{2011ApJ...730...10H}, and the Lick-Carnegie Planet Search \citep{2010ApJ...708.1366V}, are therefore increasingly focusing on very nearby, old, photometrically quiet late G, K, and M-type main sequence dwarfs. The 451 stars that constitute the HARPS Guaranteed Time Observations (GTO) Òhigh precisionÓ sample \citep{2008A&A...487..373S} encompass many of the Sun-like stars at declination, $\delta<20^{\circ}~{\rm N}$ that are currently receiving intense surveillance. 

Planets can also be detected via the wobble that they induce in their parent star's motion projected onto the plane of the sky. This so-called {\it astrometric} technique is most sensitive to massive planets orbiting stars that are relatively close to Earth. Because the star's motion is detectable in two dimensions, the planet's actual mass, rather than just the combination $M\sin i$, can potentially be measured. Planets on more distant orbits are ultimately easier to detect using astrometry because the amplitude of the star's motion is larger, but finding these planets requires a longer timeline of observations due to their greater orbital periods. To date, despite decades of optimistic expectations, see, e.g., \citet{1982ApJ...263..854B}, the astrometric method has yet to unambiguously detect an exoplanet, and the most relevant discoveries are of objects such as the 28$M_{\rm J}$ companion to nearby L1.5 dwarf DENIS-P J082303.1-491201 \citep{2013A&A...556A.133S} that lies just above the upper mass limit for planets. The drought is likely to end soon, however. Simulations by \citet{2008A&A...482..699C}, indicated that ESA's {\it Gaia} Mission, launched on December 19th, 2013, and designed to achieve 7 $\mu$as precision on $V=10$ stars, should be capable of detecting several thousand giant planets out to 3-4 AU from stars lying within 200 pc of the Sun. A more recent analysis by \citet{2014ApJ...797...14P}, which uses an updated galactic model and planet occurrence rates, and which folds in Gaia's actual on-sky astrometric performance, suggests that the mission should find $21,000\pm6,000$ long-period planets with $1M_{\rm J}<M_{\rm P}<15M_{\rm J}$ during its nominal 5-year duration, with several times more planets expected to be observed in the event that the mission is extended.

The Doppler and astrometric techniques measure a planet's gravitational pull on its star, and are thus sensitive to the planet's mass. By contrast, transit photometry detects the amount of starlight that a planet can obscure, and thus yields an estimate of the planet's size. If the line of sight from the parent star to the Earth lies in or near the orbital plane of an extrasolar planet, that planet passes in front of the disk of its star once each orbit as viewed from Earth, and periodically blocks a small fraction of the star's light. Sufficiently precise photometric time-series measurements of the star's brightness are used to reveal transits, which can be distinguished from rotationally-modulated star spots and intrinsic stellar variability by their periodicity, their effectively square-well shapes, and their relative spectral neutrality. Although geometrical considerations limit the fraction of planets detectable by this technique (planetary orbital angular momentum vectors are aligned apparently randomly with respect to the galactic plane),\footnote{The angular momentum of the Solar System ultimately derives from the turbulent motions that existed within the Sun's parent giant molecular cloud, and is not a direct consequence of the differential rotation of the galactic disk.} thousands of stars can be surveyed within the field of view of one telescope, so transit photometry is potentially quite efficient. Indeed, transit photometry now generates a clear majority of new planetary candidates, and ground-based surveys such as the HAT Net \citep{2012AJ....144...19B} and SuperWASP \citep{2013A&A...552A.120S}, are generating dozens of confirmed planets per year. 

The preliminary version of the \ik Q12 planet candidate catalog, consisting of signals identified using the first 3 years of \ik data, contains 3,538 candidate planets, at least $\sim$ 90\% of which are expected  to represent true exoplanets \citep{2011ApJ...738..170M, 2008A&A...482..699C, 2013A&A...557A.139S, 2014ApJ...784...44L}.  Almost half of these objects are associated with target stars having two or more planet candidates, implying that configurations of tightly spaced, highly co-planar planets are ubiquitous \citep{2011ApJS..197....8L}. More than 90\% of these candidates have estimated radii $R_p < 6$ R$_\oplus$ and almost half have radii $R_p < 2$~R$_\oplus$, despite observational biases against detecting smaller planets.\footnote{For an up-to-date list of \ik candidates, see \url{http://exoplanetarchive.ipac.caltech.edu/cgi-bin/ExoTables} .} In many of \ikt's multiple-transiting planet systems, the transit events exhibit deviations from strict periodicity. These so-called transit timing variations (as well as transit duration variations that are, much more rarely, also observed) can be attributed to planet-planet interactions within the systems, and have, in many cases, established to high confidence that a particular candidate system is the genuine item rather than a false positive. Scores of \ik candidates, furthermore, have been confirmed with Doppler follow-up measurements, e.g.,~\citet{2011ApJS..197....7C}, \citet{2014ApJS..210...20M}, and hundreds of candidates have been validated to be planets at very high confidence through statistical arguments \citep[e.g.][]{2011ApJ...727...24T, 2014ApJ...784...45R}. Continued follow-up observations of Kepler Objects of Interest, as well as new photometric data from the K2 Mission \citep{2014PASP..126..398H} will keep the Kepler project productive for years to come.

Reaching further into the box of astrophysical tricks, one finds the microlensing method, which has been successfully employed to investigate the distribution of faint stellar and substellar mass bodies within our galaxy \citep{1997ApJ...486..697A}, and which has now produced published detections of 18 extrasolar planets \citep{2013A&A...552A..70K}. The microlensing phenomenon stems from the general relativistic bending of the light from a distant star (the source) by a massive object (the lens) passing between the source and the observer. Such lensing can cause the source to appear to gradually brighten by a factor of a few, or for special configurations leading to high magnification events factors exceeding 100, over a time scale of weeks to months. If the lensing star has planetary companions, then these far less massive bodies can produce brief enhancements in the brightness -- generally of a few hours in duration -- provided the line of sight passes close to the planet. Under favorable circumstances, planets as small as Earth can be detected. The properties of individual planets, however, and usually even of the stars that they orbit, can only be estimated in a statistical sense because of the many parameters that influence a microlensing light curve see, e.g., \citet{1997Icar..127..269P, 2012ARA&A..50..411G}. The microlensing technique is comparatively well-suited to the detection of planets with long periods, and can readily probe Jovian, or even super-Earth mass planets in orbits similar to that of Jupiter itself. Due to the prevalence of M-type main-sequence stars in the stellar mass function, the microlensing technique has found a number of examples of  giant planets and Neptune-mass planets beyond the locations of the so-called nebular snow lines for these stars \citep{2010ApJ...713..837B}. \citet{2011Natur.473..349S} have extensively analyzed several years of microlensing survey data from the Galactic Bulge region, and report the  discovery of a population of ``isolated" planetary mass objects with $M\sim~$M$_{\rm J}$, which appear to occur with somewhat higher frequency than main-sequence stars. These planetary mass objects lie either in very distant orbits ($d>100\,{\rm AU}$) from undetected parent stars, or they are gravitationally unbound and thus floating freely through the galaxy. The recent discovery of WISE J085510.83Ð071442.5 \citep{2014ApJ...786L..18L}, a free-floating 3-10~$M_{\rm J}$ object  lying a mere 2 parsecs from the Sun, lends credence to the idea that free-floating planets could be common. Certainly, the potential presence of hundreds of billions of such drifters, should these worlds withstand statistical scrutiny, would have important consequences for the theory of planet formation, and would serve as a definitive endorsement of the microlensing technique. 

All of the foregoing are indirect methods to locate extrasolar planets, but what about imaging planets directly? While observational efforts to directly detect extrasolar planets can be traced back to Christiaan Huygens in the 1600's\footnote{Huygens, C. 1698 \textit{Cosmotheoros}, Adriaan Moetjens: The Hague, Netherlands.},  only very recently have the extreme planet-star brightness contrasts and angular separations been successfully challenged. The reflected starlight from planets with orbits and sizes like those in our Solar System is well over  twenty visual magnitudes (a factor $\gtrsim10^{8}$) fainter than the stellar brightness. Diffraction of light by telescope optics and atmospheric variability compound the difficulty of direct detection of extrasolar planets.  Technological advances, most critically in the area of adaptive optics on large ground-based telescopes and coronographic imaging from space, are enabling direct detection of relatively massive, young, and in most cases self-luminous, planets orbiting at substantial distances $a\gtrsim10\,{\rm AU}$ from young, nearby stars with masses $M_{\star} \gtrsim M_{\odot}$.  These detections leverage planet-star brightness contrasts of order $L_{\rm p}/L_{\star} < 10^{5}$ and angular separations of order $\alpha \sim 1$ arc sec. A high profile example is illustrated by  
Beta Pictoris b \citep{2010Sci...329...57L}, with mass $M_{\rm P}=8^{+5}_{-2}\, $M$_{\rm J}$ at separation $a=8^{+1.7}_{-0.4}\,$AU from Beta Pictoris, a young A-type star that has long been known to harbor an infrared excess and an extensive debris disk \citep{1984Sci...226.1421S}. Another interesting and somewhat puzzling instance is provided by Fomalhaut b \citep{2008Sci...322.1345K}. This curious object, which has been imaged with HST, lies just beyond the inner radius of the dust ring surrounding the young, nearby A-type star Fomalhaut. The imaged light associated with the planetary candidate's position is dust-scattered starlight, associated, perhaps, with a circumplanetary disk \citep{2013ApJ...769...42G, 2014arXiv1403.5268K}.

The most startling and remarkable direct-imaging detection is that of the system of four companions to the young ($\sim30\,{\rm Myr}$) A-type star HR 8799 \citep{2008Sci...322.1348M, 2010Natur.468.1080M}. These planets all have masses in the vicinity $M_{\rm P} \sim 5-7~$M$_{\rm J}$, with semi-major axes  lying at 14.5, 24, 38, and 68 AU, and call to mind a version of our own outer Solar System that has been scaled up by a factor of several in both mass and astrocentric distance. The presence of HR 8799's  planets at their observed distances presents interesting challenges to both the core-accretion \citep{lissaueretal09} and the gravitational instability \citep{2007prpl.conf..607D} models of giant planet formation. The proximity of the system configuration to the dynamical stability boundary is also intriguing, and implies either that the system will suffer catastrophe within a time that is much less than the star's main sequence life, or that the planets are dynamically protected by a resonant mechanism \citep{2010ApJ...710.1408F}. The pace of detection by direct imaging is proceeding at a rapid clip, with additional examples being those of HD 95086b \citep{2013ApJ...779L..26R}, 2MASS J01033563-5515561(AB)b \citep{2013ApJ...779L..26R}, Ksi And b \citep{2013ApJ...763L..32C}, and GJ504b \citep{2013ApJ...774...11K}, which tend to be massive planets $M\gtrsim4M_{\rm Jup}$ orbiting at $R\gtrsim40\,$AU from young (or relatively young) stars.

Direct imaging is poised for major advances with the commissioning of several new projects that utilize 8 to 10-meter class telescopes. Among these are the Gemini Planet Imager (GPI), which experienced first light at the Gemini South Telescope in late 2013 \citep{2014arXiv1403.7520M}. The GPI adaptive optics system is designed to detect planetary companions at separations of $\delta$=0.2 -1.0 arc seconds, and brightness contrasts of $\sim10^{-7}$ in 1-2 hour exposures. This observational strategy is sensitive to Jupiter-mass planets within the first several million years of their formation, and will enable direct imaging of a variety of planets with ages under a billion years. The GPI system, furthermore, uses an integral field spectrometer to obtain $R\sim40$ spectra that will give important clues to overall atmospheric compositions and states. Already, spectra of directly imaged exoplanets are giving physical information, see, e.g. \citet{2013Sci...339.1398K}, who report on the use of the Keck Telescope's OSIRIS integral field spectrograph to detect CO and ${\rm H_{2}O}$ absorption in HR 8799c. The SPHERE adaptive optics instrument on the VLT \citep{2008SPIE.7014E..18B} draws on optical and near-IR imaging, low-resolution spectroscopy, and polarimetry in the service of finding and characterizing young Jovian planets. SPHERE's overall scientific capabilities are broadly similar to those of GPI, and it has been operational since mid-2014. A third project, with similar goals, and with a high angular resolution that should permit detection of planets at $r\sim4\,{\rm AU}$, is the Subaru Coronagraphic Extreme-AO Project, \citep{2009SPIE.7440E..0OM}, which is scheduled to begin scientific operations in 2015.

Several other methods can also be used to detect extrasolar planets. The precise timing of eclipses of eclipsing binary stars has the potential of revealing the masses and orbits of unseen companions, and indeed, such timing measurements have been used to confirm the planetary nature of the transiting circumbinary planet Kepler-16 b \citep{2011Sci...333.1602D}. Spectroscopy could be used to identify gases that would be stable in planetary atmospheres but not in stars, and Doppler variations of such signals could yield planetary orbital parameters. As an example, \citet{2013ApJ...767...27B} have obtained tentative detections of ${\rm CO}$ and ${\rm H_{2}O}$ in the emission spectrum of 51 Peg during superior conjunction of its non-transiting short-period planetary companion. Starlight scattered from a planetary atmosphere will be polarized \citep[for detail, see][]{2000ApJ...540..504S}, and the expected polarization signal lies at the threshold of current detectability \citep{2008ApJ...673L..83B, 2009ApJ...696.1116W}. Radio emissions similar to those detected from Jupiter could reveal the presence of extrasolar planets, although to date, all attempts at detection via this channel have been unsuccessful, see, e.g., \citet{2010AJ....140.1929L}. Finally, the detection of artificial signals from an alien civilization would potentially imply the presence of a habitable terrestrial-sized planet or giant planet satellite.

The \ik spacecraft has broken through the psychologically important threshold of one Earth radius, identifying numerous Earth-size candidates, and has validated several planets smaller than Earth, including Kepler-37 b,  a planet about the size of Earth's Moon \citep{2013Natur.494..452B}. The
Doppler radial velocity technique is approaching the comparable 1 M$_{\oplus}$ threshold, with   the Geneva Planet Search Team's detection of an $M \sin i \sim 1.1~$M$_{\oplus}$ planet in a short-period orbit about Alpha Centauri B \citep{2012Natur.491..207D}.  All of the Earth-size exoplanets verified thus far, however, have orbital periods of  a few weeks or less, and are not worlds of genuine astrobiological interest.  Nevertheless, it has been unambiguously established that planets of masses somewhat greater than Earth's mass and larger are {\it extremely} common in the galaxy, and the emphasis of the field is shifting from the detection of extrasolar planets to their physical characterization. In the same manner that a new branch of astronomy has been created, a new and important subfield of planetary geophysics has also come into being.

\section{An Overview of the Current Planetary Catalog and Census}

Including both verified exoplanets and high-quality {\it Kepler}-detected transiting planet candidates, thousands of planets have now been identified, and the outlines of the galactic planetary census are emerging. Much of the current inventory of planets is shown in Figure 2. In this diagram, which combines planets and planet candidates from several sources, the log of the mass ratio of the planet to the central star, $\log_{10}(M_{\rm P}/M_{\star})$, is plotted on the ordinate, and $\log_{10} P$ is charted on the abscissa. For the case of Doppler detections, we conflate $M_{\rm P}$ with $M_{\rm P}\sin(i)$, and for high-quality candidate planets detected photometrically, we have transformed from planetary radius, $R_{\rm p}$ to $M_{\rm P}$ using a simple transformation
\begin{align}
M_{\rm P} &= \left(\frac{R_{\rm p}}{{\rm R}_{\oplus}} \right)^{3}{\rm M}_{\oplus} &  R_{\rm p} < {\rm R}_{\oplus} ,\\
M_{\rm P} &= \left(\frac{R_{\rm p}}{{\rm R}_{\oplus}} \right)^{2.06}{\rm M}_{\oplus} &  R_{\rm J} > R_{\rm p} > {\rm R}_{\oplus} ,\\
M_{\rm P} &= {\rm M}_{\rm J}  & \, R_{\rm J} > {\rm R}_{\oplus} .
\label{mr-relation}
\end{align}

In the above, the $M_{\rm P}/{\rm M}_{\oplus}=(R_{\rm p}/{\rm R}_{\oplus})^{2.06}$ relation \citep{2011Natur.470...53L} provides a best-fitting power-law to the Solar System planets spanned in mass by Earth and Saturn. That such a simple power law does such a good job in the face of different formation mechanisms for the solar system planets can largely be attributed to the fact that only four planets are involved in the fit, and to the substantial similarity between Uranus and Neptune. The $M_{\rm P} = $ M$_{\rm J} $ relation is applied to just a handful of candidates that appear in multiple-planet systems, and does not affect the overall trends exhibited by the diagram. In the absence of solid information, the relation $M_{\rm P} = \left({R_{\rm p}}/{{\rm R}_{\oplus}} \right)^{3}{\rm M}_{\oplus}$ simply assumes that the small number of planet candidates with radii smaller than Earth have Earth-like densities. As we discuss below, the results from planets that have had both $M_{\rm P}$ and $R_{\rm p}$ independently measured indicate that there is no single planetary mass-radius relation, and that the dispersion in radii for planets of given mass is substantially larger than was generally expected.

The degree of incompleteness that characterizes Figure 2 is somewhat difficult to evaluate, especially in light of the fact that it draws together planets detected by  a variety of methods. Nevertheless, a reasonable estimate for the current limiting sensitivity of the Doppler Surveys, in the absence of highly-focused, high-cadence campaigns on bright nearby stars, is $K=2\, {\rm m~s^{-1}}$, multiplied by an additional period-dependent term of order unity that favors shorter orbits. A curious, and mostly unpredicted feature of the diagram is that the planets fall into three  groups. Observational completeness, furthermore, while not perfectly known, see, e.g., \citet{2008PASP..120..531C}, is sufficient to ensure that the gaps separating the three distinct populations represent real features of the galactic planetary census.

Centered on $M_{\rm P}/M_{\star}\sim1\times10^{-3}$ (i.e., $\sim 1 $~M$_{\rm J}$), and $P\sim3\,{\rm days}$, is a population of {\it hot Jupiters}. The members of this group, on account of their ready detectability, are the best-studied (if still incompletely understood) population of extrasolar planets. In aggregate, hot Jupiters are clearly gas giants dominated by hydrogen -- helium compositions, but in nearly every observable property -- radii, chemistry, emission properties, density, orbital elements, and so forth, they display a tremendous variability that defies easy categorization.

Hot Jupiters constitute a prominent fraction of the planets in Figure 1 because they are readily detectable, but as a whole, they represent a small fraction of the entire population
of planets. Among a large sample of main-sequence host stars, they are the exception rather than the rule, with an occurrence rate among nearby Sun-like main-sequence stars of only about $\sim$1\% \citep{2011AN....332..429M, 2012ApJ...753..160W}. From the \citet{2013ApJS..204...24B}
catalog of \ik candidate planets, one estimates a hot Jupiter occurrence rate of 0.5\%, in contrast to the
order-unity occurrence rate of smaller planets with orbital periods $P<100\,{\rm d}$.

\begin{figure}[H]
\centering
\includegraphics[height=14cm,angle=0]{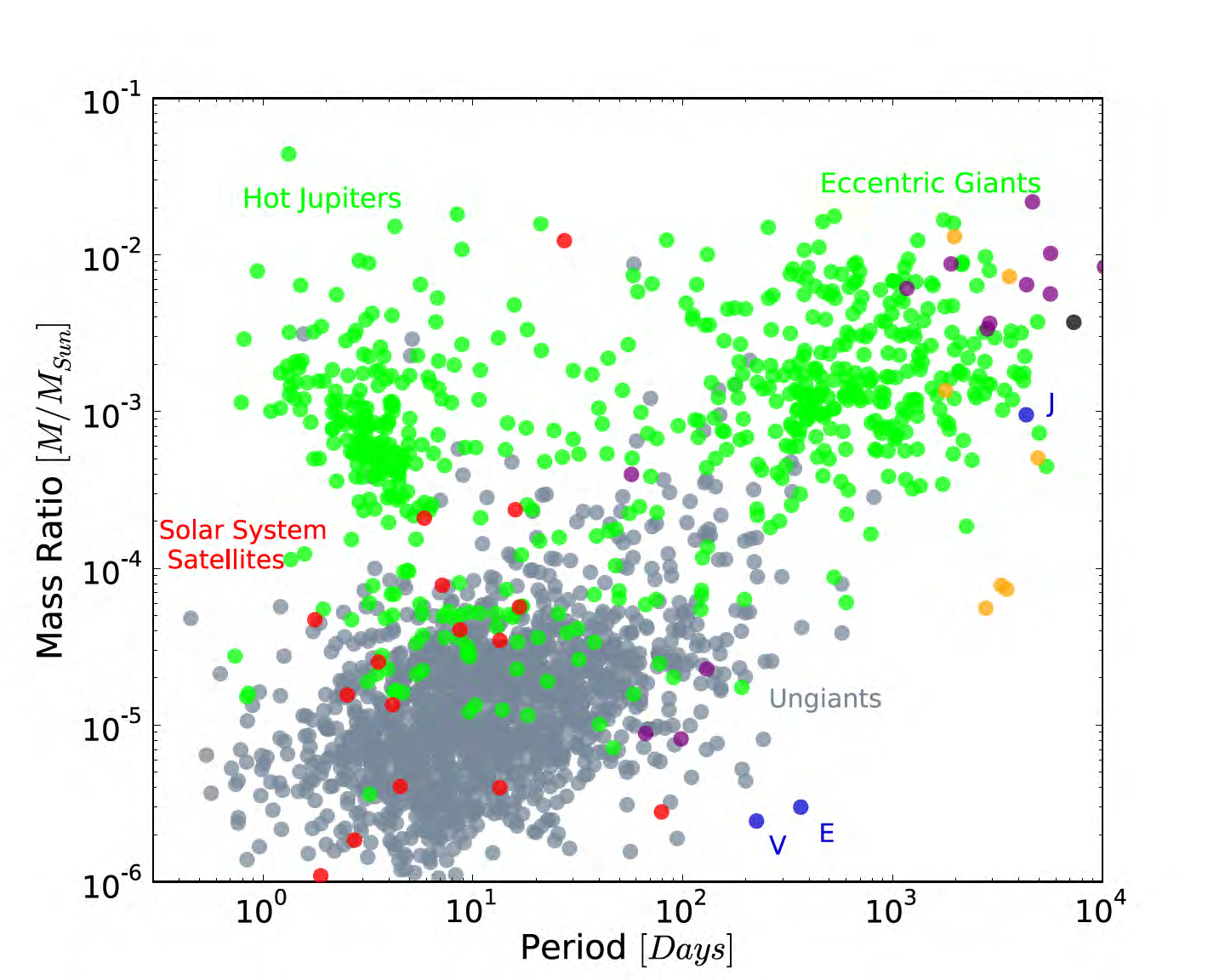}
\caption{ {\it Green circles:} $\log_{10}(M_{\rm satellite}/M_{\rm primary})$ and $\log_{10}(P)$ for 634 planets securely detected by the radial velocity method, including those first identified via transits. {\it Red circles:} $\log_{10}(M_{\rm satellite}/M_{\rm primary})$ and $\log_{10}(P)$ for the regular satellites of the Jovian planets in the Solar System, along with Earth's Moon.  {\it Gray circles:} $\log_{10}(M_{\rm satellite}/M_{\rm primary})$ and $\log_{10}(P)$ for 1501 \ik candidates and objects of interest in which {\it multiple} transiting candidate planets are associated with a single primary. Radii, as reported in \citet{2013ApJS..204...24B}, are converted to masses using the mass-radius relations described in the text. {\it Black circles:} Planets detected by direct imaging, with mass ratios estimated from infrared emission detected from the planetary candidate (imaged planets, e.g. HR 8799~b, c, d, \& e with $P>10^{4}$~d are not shown), {\it Gold circles:} Planets detected via microlensing, {\it Purple circles:} Planets detected via pulsar and white dwarf timing. {\it Blue circles:} Jupiter, Earth and Venus.}
\end{figure}

The upper-right hand corner of Figure 2 shows a second population of planets that have been revealed by the Doppler velocity surveys. In the absence of a consensus terminology, we refer to this group as  the {\it Eccentric Giants}. Planets in this group are characterized by (1) orbital periods in the range of hundreds to thousands of days, (2) masses that are, on average, several times larger than the mass of Jupiter, with a median value of $M\sin i=1.94~$M$_{\rm J}$, although the mean of the underlying distribution is clearly smaller because more massive planets are easier to detect, and (3) typically substantial eccentricities, with a median value $e=0.22$, and $e>0.5$ not uncommon, although a few members travel on nearly circular orbits. The aforementioned HD 114762~b \citep{1989Natur.339...38L}, with $M\sin i=11\, $M$_{\rm J}$, and $P=114$~d is a member of this population. Of order ${\cal F}\sim5$\% of the solar-type F, G \& K main-sequence dwarfs in the local galactic neighborhood harbor an eccentric giant, making them roughly ten times more common than hot Jupiters. Like the hot Jupiters, however, the occurrence rate of eccentric giants depends strongly on the metallicity of the parent star \citep{1997MNRAS.285..403G, 1999MNRAS.308..447G}. This strong correlation imparts an important, albeit still imperfectly understood, clue regarding the nature of the planet-forming process. The most metal-rich stars, with $[{\rm Fe}/{\rm H}]>0.3$, i.e., more than twice the solar value, are a factor of five to ten times more likely to harbor a hot Jupiter or an eccentric giant than are stars with half a solar metallicity (i.e., those with $[{\rm Fe}/{\rm H}]<-0.3$) \citep{2001A&A...373.1019S, 2005ApJ...622.1102F}. Stars with $M_{\star}>$ M$_{\odot}$ are also statistically more likely to harbor a hot Jupiter or an eccentric giant. 

Jupiter lies near the  long-period edge of the currently known population of eccentric giants. Until recently, this placement near the tail of the distribution was attributed to observational incompleteness among genuinely jovian analogs with Jupiter-like masses and Jupiter-like periods. That is, the apparent dearth of planets with $P\sim10\,{\rm years}$ and $M\lesssim \, $M$_{\rm J}$ was thought to be the simple consequence of long required time scales for observation, and the difficulty in detecting low-amplitude signals. Such assumptions, however, are increasingly being called into question. As an example, consider the stars in the Keck radial velocity survey as maintained by the Lick-Carnegie Exoplanet Search \citep{2010ApJ...708.1366V}. By the end of 2012, the Keck I telescope had been used to obtain 31,910 Doppler measurements in the overall radial velocity database (binned to a minimum cadence of two hours in the frequently occurring case of multiple back-to-back exposures that average over stellar p-modes) on 1,298 individual, primarily solar-type, main sequence dwarfs. The median derived instrumental uncertainty on these measurements was $\sigma=1.45\, {\rm m s^{-1}}$. Stellar jitter, depending on the star, generally contributes $\sigma=1-5\, {\rm m s^{-1}}$ in quadrature to the instrumental uncertainty, and for observations associated with individual stars, the median time base line was ${\rm HJD}_{\rm obs. last}-{\rm HJD}_{\rm obs. first}$= 2,426 days. Furthermore, there were 237 stars with time baselines longer than Jupiter's 4333 d orbital period. The stars in this long-baseline group tend to be similar to the Sun in terms of basic properties, and are generally chromospherically quiet and photometrically stable, averaging 55 Doppler measurements each. A true-Jupiter analog ($P=11.862\,$ years, $K=12\,(\sin i)\,{\rm m~s^{-1}}$) orbiting almost any  of these stars would be immediately evident.

Of potentially equal significance to the concentrations of planets in the mass -- period-ratio diagram are the regions of parameter space where the number density of worlds is low. There is some evidence that Nature tends to avoid siting giant planets with orbital periods in the $10\,{\rm d}<P<100\,{\rm d}$ range; the underlying cause of these broad areas of lower concentration is not fully understood. There appears to be a general dearth of planets of both short and intermediate period with masses ranging from roughly 40 M$_{\oplus}$ to 100 M$_{\oplus}$, consistent with the predictions of rapid growth through this mass range in the core-nucleated accretion model of giant planet formation \citep{pollacketal96}.

Perhaps the most important exoplanet-related discovery of the past ten years has been the realization that of order half of the Sun-like stars in the solar neighborhood are accompanied by systems of one or more planets with periods ranging from days to months, and masses falling in the $1\,$M$_{\oplus} < M_{\rm P} < 50\, $M$_{\oplus}$ range \citep{2009A&A...493..639M, 2011arXiv1109.2497M}. These planets are clearly visible as a third population that is distinct from the hot Jupiters and from the eccentric giants. Its members constitute the bulk of the transiting planet candidates in multiple-transit systems detected by \textit{Kepler}, which are indicated in gray on the mass -- period-ratio diagram, and they tend to have relatively low orbital eccentricities.  There is clearly a desert (a dearth of objects), between giant planets, and these ``ungiants'' -- the term of art that we will employ for what is largely still terra incognito -- the planetary mass range M$_{\oplus}<M_{\rm P}< 30~{\rm M}_{\oplus}$.  

Whereas the period distribution of giant planets reveals that hot Jupiters are a separate population, distinct from the more distant giants, whose abundance peters out at periods of less than a few months, no such distinction is seen among the ungiants.  The population of these planets appears roughly constant in log period space (as would be expected from stability limits imposed by planetary dynamics) for orbital periods between 10 and 100 days.  Data are too sparse to extend this conclusion to longer periods.  At shorter periods, there is clearly a drop in the abundance of ungiant planets, but this is gradual, with no counterpart to the spike in the 3-4 day range seen for giants.  Neptune-size planets are detected down to orbital periods of 2 days, and planets with $R_p \lesssim 2$ R$_\oplus$ are observed at periods well under one day.  The current record-holder for ultra-short period planets is Kepler-78 b, a refractory world (it has been detected both in transit and using RV) slightly larger than the Earth with an orbital period of only 8.52 hours \citep{2013ApJ...774...54S}.

Little is known about the physical properties of ungiants, apart from that \emph{by mass} they are composed primarily of elements heavier than helium, in contrast to giant planets, and that they are a diverse population in terms of dominant constituent \emph{by volume}.   It is not clear, for example, whether most of its members are properly described as ``super-Earths'' or ``sub-Neptunes'' or something else entirely. Curiously, our own Solar System, which contains nothing interior to Mercury's 88-day orbit, lacks any of the larger and closer-in planets that dominate this part of the exoplanet diagram; how common terrestrial planet configurations similar to our Solar System are is unknown, although many stars clearly host different and easier-to-detect systems of ungiants.

\newpage
\begin{figure}[H]
\includegraphics[height=14.0cm,angle=0]{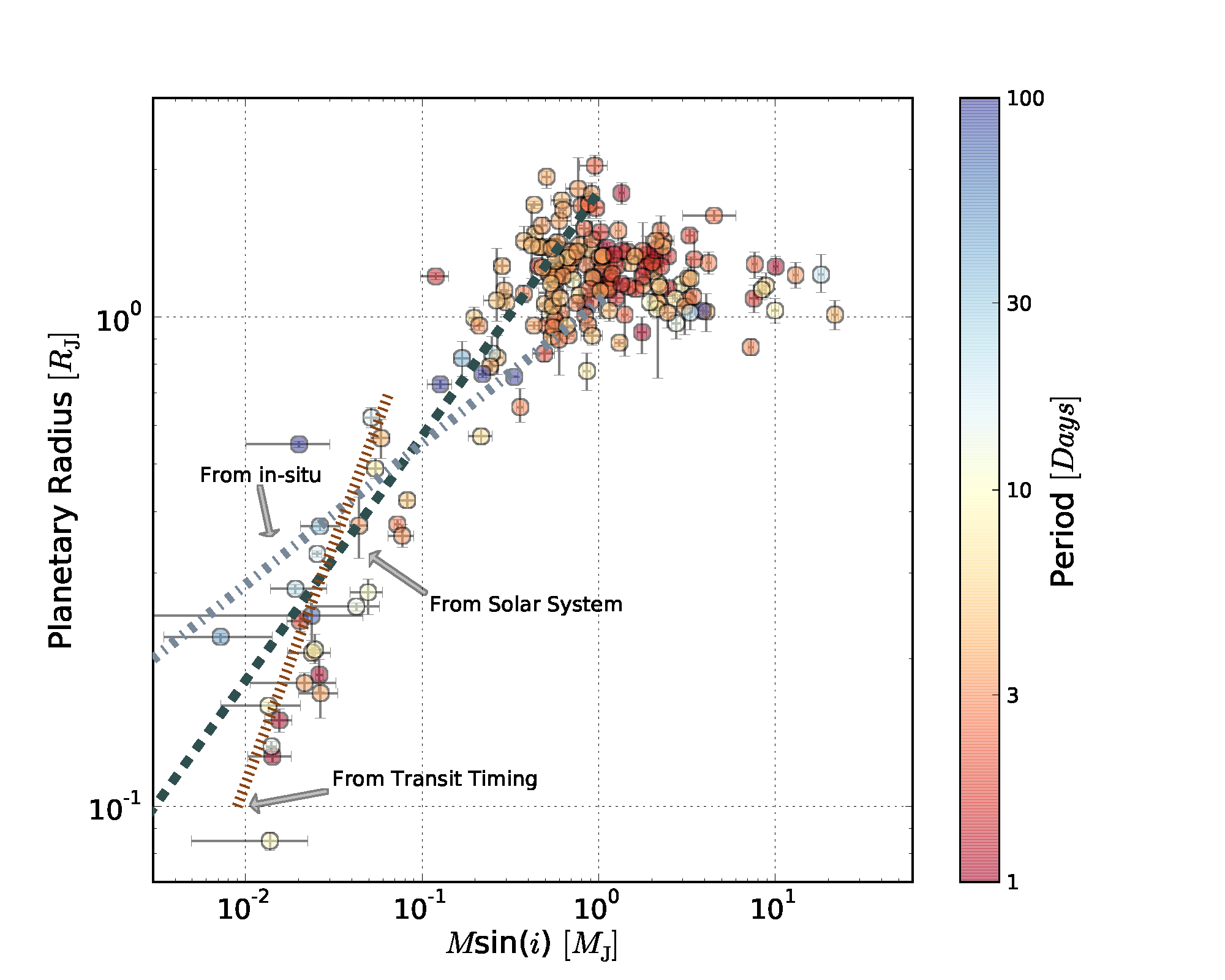}
\caption{ Mass-Radius diagram for Extrasolar planets. This figure shows planets for which both mass and radius estimates exist, with orbital period coded to run from one day (blue) through 100 days (red). Three sample mass-radius relations are shown. The short-dotted line is drawn from a simple in-situ formation theory as described in \citet{2013MNRAS.431.3444C}, and is based on rock-iron planets surrounded by hydrogen-helium envelopes, and does not include the possible effects of insolation-driven atmospheric mass loss. The long-dashed line is the best-fit $M_{\rm P}/$M$_{\oplus}=(R_{\rm p}/$R$_{\oplus})^{2.06}$ power-law appropriate to the Solar System planets spanning masses from Earth through Saturn. The medium-dashed line is the $3 M_{\rm P}/$M$_{\oplus}=R_{\rm p}/$R$_{\oplus}$ relation derived by \citet{2012ApJ...761..122L} from consideration of the transit timing variations observed by \ik for 22 multiply-transiting candidate planet pairs in the vicinity of mean motion resonances.}
\label{fig:MassRadiusDiagram}
\end{figure}
\newpage

A confounding factor is presented by the difficulty in obtaining precise Doppler radial velocity-based or transit timing-based mass estimates for 
super-Earth/sub-Neptune class planets. Figure  \ref{fig:MassRadiusDiagram} charts a selection of planets for which both mass and radius (and hence density) measurements have been made. Even with the log-log scaling, the plot indicates to the eye that planets are subject to a wide variety of structures. For example, to within the errors, it appears that plants of $M\sim6~$M$_{\oplus}$ vary in radii by a factor of at least three. When mass is viewed as a function of radius, the transition in composition appears sharper, with planets $R_{\rm p}\gtrsim 2$ R$_\oplus$ generally not dominated by rock (including iron), and those with radii $R_{\rm P}\lesssim 1.6$ R$_\oplus$ and measured masses mostly rocky.  While the upper limit to the size of abundant rocky planets within $a\sim 0.5$ AU of their stars seems well-established, it is quite possible that many or even most planets smaller than 1.6 R$_\oplus$ are composed primarily of low density constituents, but that the low masses of such planets makes them difficult to measure.

On Figure \ref{fig:MassRadiusDiagram}, we have over-plotted three potential mass-radius relations for ungiants. The first relation is the $M_{\rm P}/$M$_{\oplus}=(R_{\rm p}/$R$_{\oplus})^{2.06}$ best-fit to the Solar System planets spanned in mass by Earth and Saturn. Despite the uncertainties, and the clear intrinsic variations, this relation provides a reasonable fit to the mass-radius relation of planets smaller in mass than Saturn. The Solar System fit is bracketed by two alternate relations. The first, with a steeper slope, is derived from the assumption that low-mass, short-period planets have rocky cores surrounded by hydrogen-helium envelopes, with envelope masses dictated by {\it in situ} formation, as discussed in \citet{2013MNRAS.431.3444C}. The second alternate relation, $3M_{\rm P}/$M$_{\oplus}=R_{\rm p}/$R$_{\oplus}$, has a shallower slope, and is a fit derived from an ensemble of transit timing measurements applied to 22 systems within \ikt's multi-transit transiting planet candidate population by \citet{2012ApJ...761..122L}. Recent work by \citet{2014ApJ...783L...6W} has generated additional Doppler velocity-based mass measurements for a selection of \ik candidates that are not shown in Figure \ref{fig:MassRadiusDiagram}. These latest results reinforce the conclusion that planets with $R_{\rm p}<4R_{\oplus}$ span a very wide range of masses.

 \ik  is capable of detecting rocky terrestrial planets analogous to Venus and Earth, but  the spacecraft is unlikely to find a statistically robust sample of terrestrial planets similar to those in our Solar System. The mass-period diagram is thus still largely incomplete for terrestrial planets with  Solar System-like masses and orbits, and will likely remain so for some time. In the absence of solid data, we can only resort (with an abundance of cautionary disclaimers) to statistical extrapolations to anticipate the inventory of low-mass planets.

\begin{figure}[ht]
\centering
\includegraphics[height=9.5cm,angle=0]{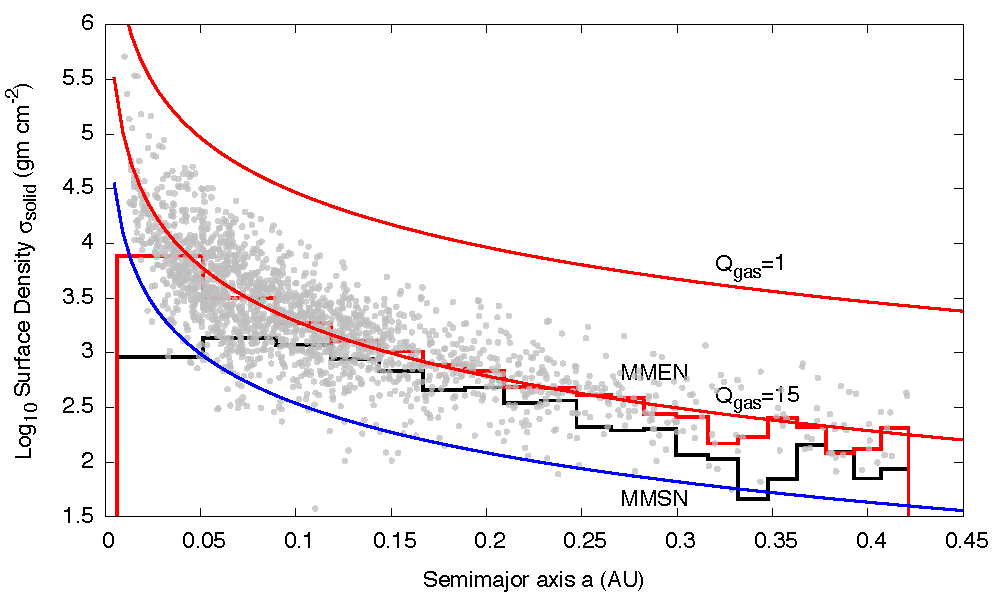}
\caption{ The solid surface density profile of the ``Minimum-Mass
  Extrasolar Nebula'' (MMEN) constructed from {\it Kepler} data \citet{2013MNRAS.431.3444C}.  {\it Gray circles:} $\sigma_{{\rm solid},i} \equiv M_i/2\pi a_i^2$, computed from {\it Kepler} planets with $R_p < 5$ R$_\oplus$ and $P < 100$ d, assuming $M_p/$M$_\oplus = (R_p/$R$_\oplus)^{2.06}$ and solar-mass host stars. For reference, $\sigma_{\rm solid}$ for the minimum-mass Solar nebula (MMSN) --- or technically its extrapolation inward, since no planet is present interior to Mercury at $a = 0.4$ AU --- is plotted as a blue solid curve. As plotted, the MMEN is a factor of 5 times more massive than the MMSN at these distances. $Q_{\rm gas} <1$, where $Q$ is the Toomre stability criterion \citep{toomre64},  implies the nebular surface density in gas required for axisymmetric gravitational instability.}
\label{fig:MMENfromKepler}
\end{figure}

To construct a concrete (albeit uncertain) extrapolation, we assume that planet formation is highly efficient, and that long-distance migration ($\Delta a \gtrsim a$), where $a$ is the current semi-major axis of the planet and $\Delta a$ the amount it has migrated, in the super-Earth population does not regularly occur. We then idealize the population of \ik planetary candidates from \citet{2013ApJS..204...24B} as being composed of solar-composition solids, and we assign each planet a surface density, $\sigma_{\rm solid}=M_{\rm P}/2\pi a^{2}_{\rm p}$, where $a_{\rm p}$ is the orbital semi-major axis, computed with the approximation that $M_{\star}=M_{\odot}$, and $M_{\rm P}=(R_{\rm p}/$R$_{\oplus})^{2.06}\,$M$_{\oplus}$. The resulting $N$=1,925 surface densities obtained for low-mass candidates with $R<5~$R$_{\oplus}$ are plotted in Figure \ref{fig:MMENfromKepler}. The best-fit power law to the median data can be interpreted to define a ``minimum mass extrasolar nebula''
$\sigma_{\rm solid}=50 \left({a}/{1\,{\rm AU}}\right)^{-1.6} {\rm g\, cm^{-2} }$,
which has a very similar power-law index, $\alpha=-1.6$ to the canonical minimum mass solar nebula (MMSN) value of $\alpha=-1.5$ \citep{1977MNRAS.180...57W}, but with a density normalization that is $5\times$ larger. 

Extending the $\sigma_{\rm solid}=50 \left({a}/{1\,{\rm AU}}\right)^{-1.6} {\rm g\, cm^{-2} }$ power law to a distance $a\sim1\,{\rm AU}$ leads to two suggestions, or better, to two moderately informed speculations: 
\begin{itemize}
\item
Roughly half of the sunlike stars in the Sun's neighborhood will harbor a planet as large as (or larger than) Earth in (or near) an Earth-like orbit.
\item
Among the planetary systems of single main-sequence stars, the inventory of mass in the Solar System's terrestrial planets will lie below the median total planetary mass interior to $a\sim~2~$AU.
\end{itemize}

\section{Planetary Geophysics}
Investigations of extrasolar planets can be divided conceptually into (1) analyses of planetary structure and composition, (2) the exploration of planetary orbital dynamics, and (3) the study of planet formation. These three areas closely inform one another. For example, at a given planetary age, $\tau_p$, the radius of a giant planet (a zeroth-order structural property) depends on the mode of formation. A planet that emerges from the process of core accretion will probably be smaller at early $\tau_{\rm p}$ than one that arises from disk gravitational instability \citep{2007ApJ...655..541M}. The planetary radius, in turn, has a strong influence on the tidal dissipation rate, with tidal luminosity obeying $\dot{E}\propto(R_{\rm p}/a)^{5}$, and frequently, the long-term dynamical behavior of a compact planetary system is sensitive to its tidal evolution, e.g., \citet{2002ApJ...573..829M}. In the remainder of this article, we will give an overview of the ``geophysics'' of extrasolar planets, with an eye toward elucidating the connections to theories of dynamical evolution and theories of planetary formation.

The direct characterization of extrasolar planets has been enabled in large part by elaborations on transit photometry, and as a result, our understanding is strongly biased toward short-period planets orbiting close to their parent stars.  Most of the well-characterized giant exoplanets were discovered by ground-based transit searches, with most of the ungiants found by the Kepler spacecraft \citep{2014Natur.513..336L}.  With this context, we then provide a synopsis of the theory of planet formation, largely informed by the contents of our own Solar System. 

\subsection{Exoplanetary Systems}
Radial velocity surveys have detected scores of systems with more than one exoplanet, including more than one dozen nearby stars with three or more planets. In systems with high multiplicity, the planets tend to have low masse and low eccentricities, although a number of systems with three or more giant planets are known.  Planets tend to be clustered into systems in the sense that there are more multiple planet systems than would be the case were planets randomly distributed among stars.  Hot Jupiters are an exception to this rule -- few of them are known to have companion planets. (Nonetheless, the star $\upsilon$ Andromedae has at least three jovian-mass planets, one of which is a
hot Jupiter. The remaining planets are far more distant
from $\upsilon$ And and travel on eccentric orbits.

The multi-transiting systems from \ik provide a much larger and very rich dataset of short-period (primarily from a few days to several months) planets that can be used to powerfully test theoretical predictions
of the formation and evolution of planetary systems.

The major findings, updated from \citep{2011ApJS..197....8L} are:
\begin{enumerate}

\item The large number of candidate multiple transiting planet systems observed by \ik show that flat (i.e., nearly coplanar) multi-planet systems are
common
in short-period orbits around other stars.  This result holds for ungiant planets in the size range of $R_{\rm p} \sim$ 1 -- 6 R$_\oplus$, but not for giant planets.
Not enough data are yet available to assess its viability for smaller worlds, nor for planets with orbital periods longer than a few
months.

\item Mean-motion orbital resonances and near resonances are mildly statistically favored over random period ratios, but most
planets are neither in nor very near such resonances. First-order resonances (e.g., period ratios of 2:1, 3:2, etc.) dominate, but second-order resonances (with period ratios 3:1, 5:3, etc.) also are
 manifested. Most near-resonant planet pairs  have period ratios from one to a few percent larger than those of the nearby resonance. Few planet pairs  have  period ratios slightly smaller than those of first-order resonances. 

\item Almost all candidate systems survive long-term dynamical integrations that assume circular, planar orbits and the mass-radius
relationship given in equations (2) - (4), derived from planets within our Solar System.

\item Attempts to generate a simulated ensemble of planetary systems to match the observed \ik planets do not strongly prefer fits by a single homogeneous population. In
particular, there is some evidence for a population or subpopulation of systems that either contains only one detectable planet per star or multiple
planets with high relative orbital inclinations that produces the observed excess of singly-transiting systems. A single population, whose
characteristics are described below, can account for the vast majority of multi-transiting systems, as well as roughly half of
the single planets observed. A third, rarer, group of planar densely-packed multi-planet systems, also appears to be present. Note that these populations of planetary systems need not be cleanly separated, i.e., there may also be significant numbers of intermediate systems.

\item The abundance and distribution of multiply transiting systems implies that the mean number of planets in the 1.5 R$_\oplus < R_{\rm p} < 6 $ R$_\oplus$
and $3 < P < 125$ day range per star for those stars with at least two such planets of this radius is $\sim$3.  According to \ik observations, only
$\sim$5\% of stars have planetary systems within these size-period limits.

\item The inclination dispersion of most candidate systems with two or more
transiting planets appears to have a median of $\sim 2^{\circ}$ (e.g., \citealt{2011ApJS..197....8L}), suggesting relatively low mutual
inclinations similar to the Solar System.

\item Many \ik targets with 1 or 2 identified planet candidates must be multi-planet systems where additional planets are present that are  not transiting and/or too small to be detected.
\end{enumerate}

\subsection{The Radii, Densities, and Compositional Structures of Extrasolar Planets}

Hot jupiters, which have $P\lesssim7$ d and $M_{\rm P}\gtrsim100\,{\rm M}_{\oplus}$, represent extrasolar planets at their most alien, but they are nevertheless by far the best studied. The short orbital periods and large masses of these worlds allow for precise and full orbital characterization for cases in which transits occur,\footnote{Transit probabilities are given by ${\cal P}_{tr}=0.0045\left(\frac{1\,{\rm AU}}{a}\right)\left(\frac{R_{\star}+R_{\rm p}}{{\rm R}_{\odot}}\right)\left[\frac{1+e\cos(\pi/2-\omega)}{1-e^2}\right]$, where $\omega$ is the longitude of periastron of the star such that $\omega=90^{\circ}$ corresponds to the prospective transit midpoint, and $e$ is the orbital eccentricity. ${\cal P}\sim 10\%$ for typical hot Jupiters. For a distant observer sited at a random viewing angle, ${\cal P}_{\oplus}=0.5$\%.} and their high surface temperatures generate infrared planetary fluxes that are readily detectable. Scores of short-period giant planets now have accurate physical measurements, but the spectra and even the radii of these planets have defied straightforward explanation. The origins, furthermore, of the majority of the physically characterizable extrasolar planets are not yet fully clear. It is generally thought that conditions at $a\lesssim 0.1 - 1\,$AU in protostellar disks are not favorable to the successful completion of the core accretion process \citep{Bodenheimeretal00}. It is therefore believed that planetary migration, either via quiescent disk processes \citep{LBR06, 2011exop.book..347L}, via planet-planet scattering \citep{2008ApJ...686..603J} or via Lidov-Kozai Cycling with Tidal Friction \citep{2001ApJ...562.1012E, FabTrem07}, or some combination thereof, plays an important role in delivering the planets to their current locations.

\begin{figure}[ht]
\includegraphics[height=10.5cm,angle=0]{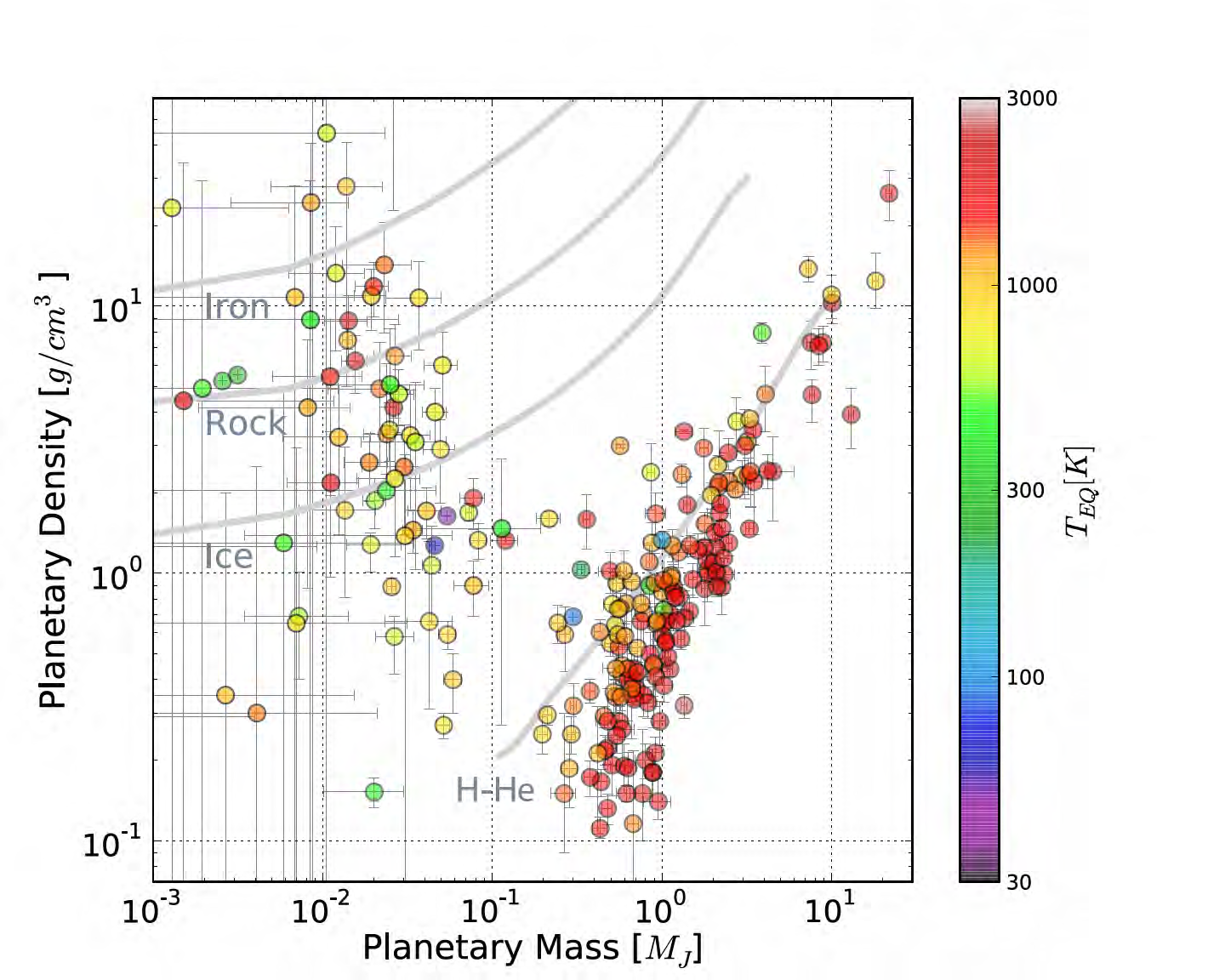}
\caption{Densities and masses for planets that have individually measured values for $R_{\rm p}$ and $M_{\rm P}$. Theoretical mass-radius relations for planets of various pure compositions are shown for comparison. Note that many of the measurements for low-mass planets were derived from transit timing variations, and these data points tend to have large uncertainties. The six most  massive Solar System planets (Venus, Earth, Uranus, Neptune, Saturn, and Jupiter) are included on the diagram, and can be distinguished on account of their low effective temperatures. }
\label{fig:MassesVsDensities}
\end{figure}

The characterization of extrasolar planets with $M_p<30$ M$_{\oplus}$ is difficult, even when the orbital periods are short. Low mass planets have small transit depths, small radial velocity half-amplitudes, and generally small transit timing variations, which make it hard to accurately determine densities. Nevertheless, dedicated follow-up on small-radius transiting planets identified by the \ik Mission has begun to yield results \citep{2014ApJ...783L...6W, 2014ApJ...785...15J}, and Figure \ref{fig:MassesVsDensities} shows that the broad outlines of the planetary distribution for M$_{\oplus}<M_{\rm P}<30$ M$_{\oplus}$ are being delineated. While this diagram is generally consistent with the $\rho(R_{\rm p})$ values exhibited by the Solar System planets, it is clear that the ungiant regime spanning (M$_{\oplus}< M_{\rm P}<30$ M$_{\oplus}$) encompasses a very wide range of planetary densities (and by extension, planetary compositions). Frustratingly, there is substantial  compositional degeneracy if planetary mass and radius are the only quantities that are measured. Often, a substantial range of  admixtures of metallic, rocky, icy, and enveloping gaseous components can jointly contribute to produce a planet of given mass and radius \citep{2008ApJ...673.1160A}. The lack of significant concrete physical data for low-mass planets means that guidance must be drawn from theoretical work. The internal structures of planets more massive than Earth are reviewed in \citet{2014arXiv1401.4738B} and the references therein. The expected properties of ``ocean'' planets with large water fractions have been discussed by a number of authors, see, e.g., \citep{2007Icar..191..337S, 2011exop.book..375S}.

Figure \ref{fig:MassesVsDensities} suggests that the overall distribution of  planetary densities reaches a broad minimum in the region $\rho_{\rm min}\sim0.2 \rightarrow 1.0\, {\rm g\,cm^{-3}}$, at roughly $M_{\rm P}\sim 0.1\,{\rm M}_{\rm J}$. Above this mass, planetary compositions are dominated by hydrogen and helium. Over 130 well-characterized examples of transiting H-He dominated giant planets (with $M_{\rm P}/{\rm M}_{\oplus} \gtrsim 100$) are now known. It is clear that within this higher mass regime, variations in mass and stellar insolation are not the only factors responsible for the observed range in planetary sizes, and {\it radius anomalies} --  observed planetary sizes that are substantially different from those predicted by structural models --  have been evident since the discovery of the first transiting extrasolar planets. Early structural evolutionary models for giant planets computed by, e.g., \cite{Bodenheimer01} and \cite{Guillot02}, suggested that a H-He dominated planet with HD 209458b's mass, insolation, and age should have a radius of approximately $R_{\rm p}\approx~1.1{\rm R}_{\rm J}$, a figure that is startlingly at odds with the observed value, $R_{\rm p}=1.38\pm0.02\,{\rm R}_{\rm J}$ \citep{Southworth10}.

\begin{figure}[ht]
\includegraphics[height=11.5cm,angle=0]{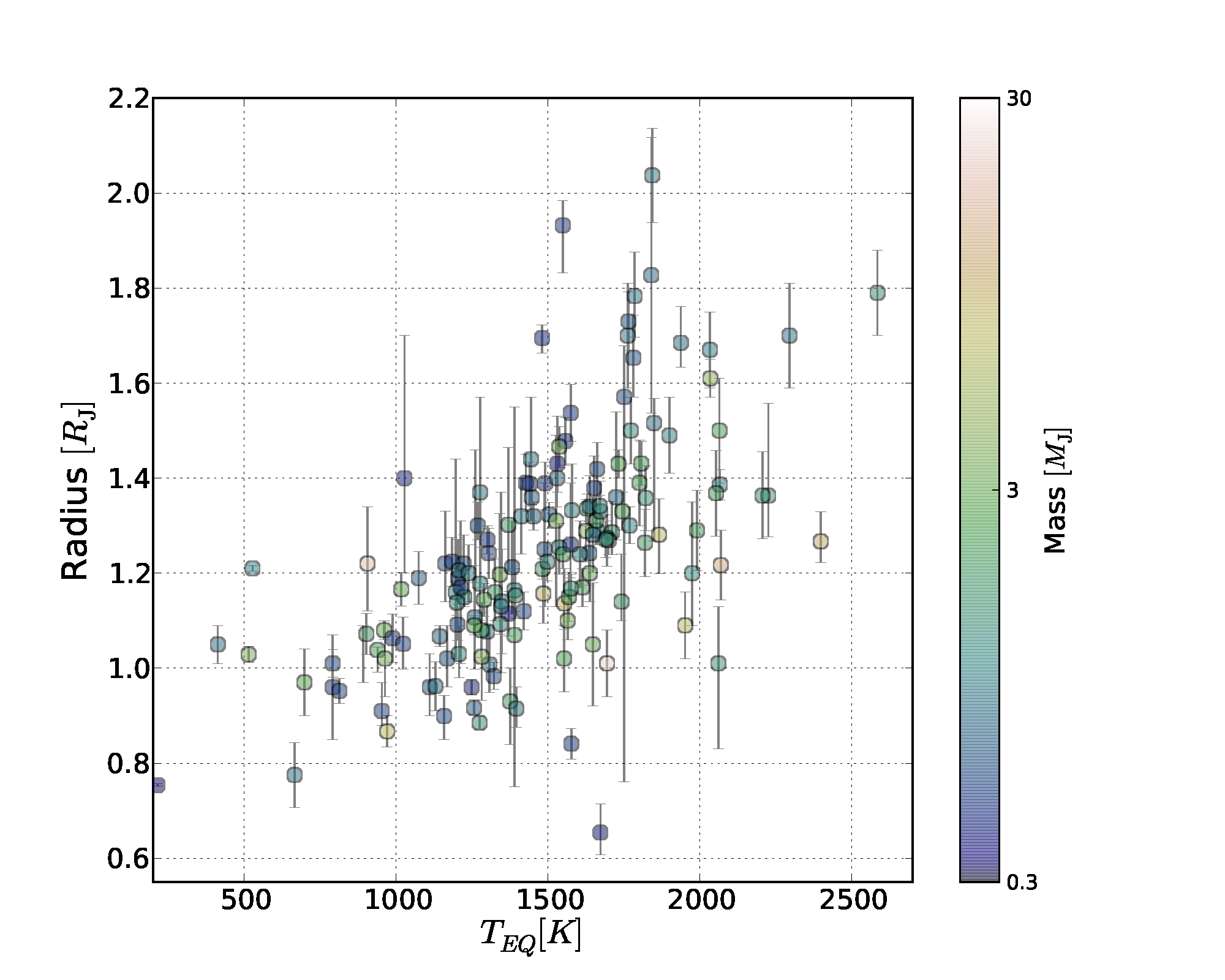}
\caption{ Planetary radii, $R_{\rm p}/{\rm R}_{\rm J}$, versus insolation-derived planetary equilibrium temperatures, $T_{\rm eq}$ (see Equation \ref{TEQn}), for 130 planets with $M_{\rm P}/$M$_{\oplus}>100$ that have both transit and radial velocity characterizations. The planets are color-coded by mass, suggesting that planets substantially more massive than Jupiter are less likely to exhibit significantly inflated radii.}
\label{fig:Giants_RvsT}
\end{figure}

During the last decade, the accumulating measurements of giant planets have demonstrated that radius anomalies are by no means anomalous, see, e.g., \cite{BurrowsOrton10}, or \cite{Baraffe10} for overviews. The substantial variation in observed radii for the giant planet population is indicated by Figure~\ref{fig:Giants_RvsT}, which charts the measured radii and uncertainties for the transiting giant planets with accurately measured masses against an insolation-based orbit-averaged planetary equilibrium temperature given by 
\begin{equation}
T_{\rm eq}={\left(\frac{R_{\star}}{2a}\right)}^{1/2} \frac{T_{\star}}{(1-e^{2})^{1/8}}\, ,
\label{TEQn}
\end{equation}
which implicitly assumes that the planet has an albedo of zero and re-radiates energy uniformly over its entire surface and depends on the average flux received by the planet over its orbit. Incorporating the effect of albedo results in multiplying the right-hand side by $(1-A)^{1/4}$. In Equation \ref{TEQn}, $T_{\star}$ is the stellar effective temperature, $e$ is the planetary orbital eccentricity, $R_{\star}$ is the parent star's radius, and $a$ is the planetary semi-major axis. The actual temperatures in most of the exoplanetary atmospheres are unmeasured, and, depending on the atmospheric properties, chemistry and dynamics, may be higher or lower than $T_{eq}$. 

\begin{figure}[ht]
\centering
\includegraphics[height=11.5cm,angle=0]{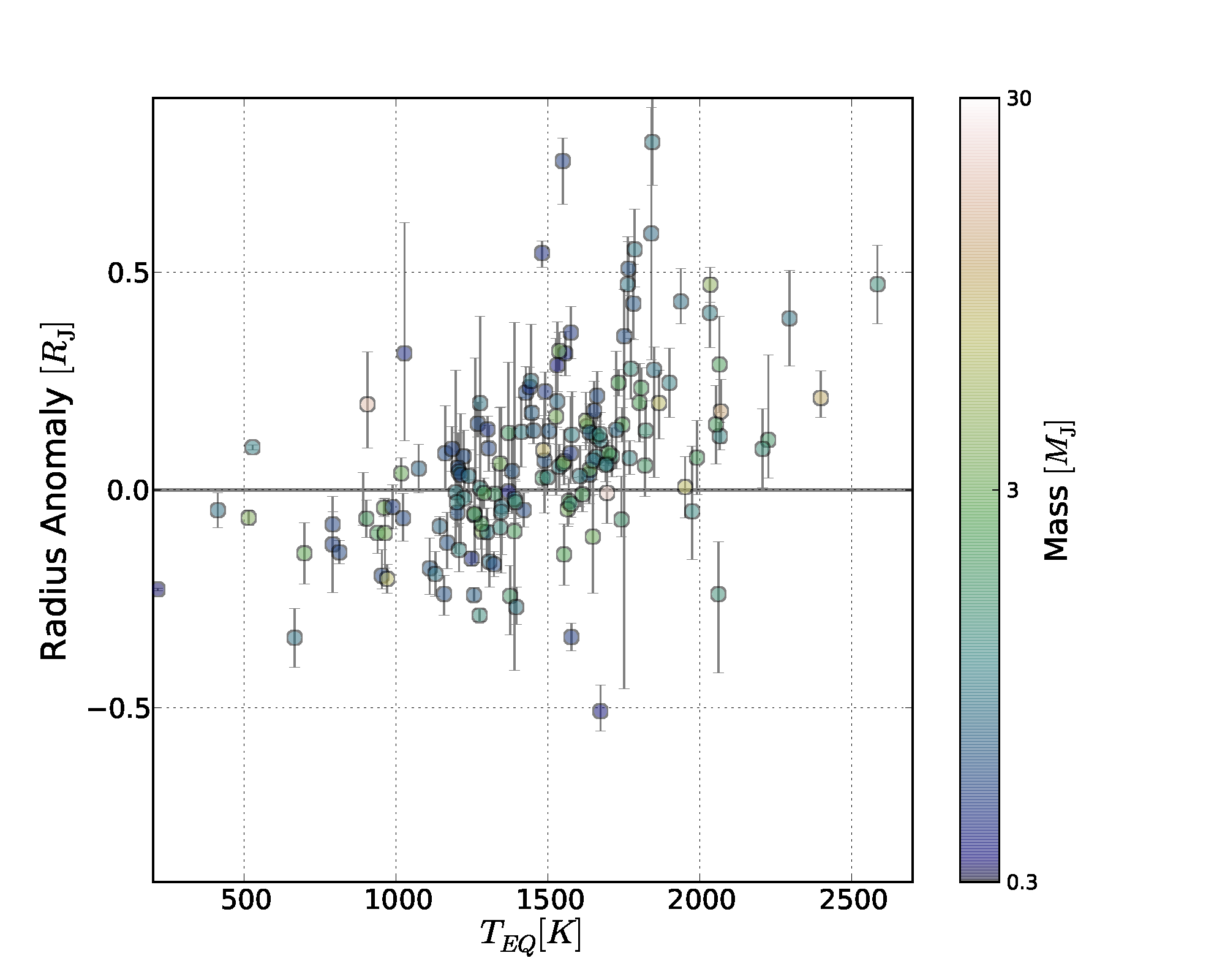}
\caption{ ``Radius anomalies'', ${\cal R}=R_{\rm obs}-R_{\rm pred}$, versus insolation-based planetary equilibrium temperature, $T_{\rm eq}$. Planetary models which account for insolation typically predict radii that are too small when $T_{\rm eq}>1200\,{\rm K}$, and too large when $T_{\rm eq}<1000\,{\rm K}$.}
\label{fig:Giants_AnomsvsT}
\end{figure}

At a given planetary mass, it is expected that the radius of a mature gas-giant planet should primarily be determined by the amount of radiative energy that it receives from its star. This expectation is supported by the clear trend in Figure~\ref{fig:Giants_RvsT}, which indicates that $R_{\rm p}$ is positively correlated with planetary $T_{\rm eq}$. The degree of quantitative agreement can be obtained by computing model radii, $R_{m}$, for H-He composition planets as a function of $M_{\rm P}$ and $T_{\rm eq}$, and comparing with those of the observed planets. We define the radius anomaly, ${\cal R}$ as the difference ${\cal R}=R_{\rm obs}-R_{\rm pred}$ between the observed and predicted radii. Figure \ref{fig:Giants_AnomsvsT} shows the result of such a comparison to a set of baseline core-free models computed by \citet{2003ApJ...592..555B} using a Henyey-type planetary structure and evolution code. 

Figure~\ref{fig:Giants_AnomsvsT} indicates that the structural models do explain some, but not all, of the observed variation in planetary radii. For $T_{\rm eq}\gtrsim1000$ K, planets are usually larger (and sometimes substantially larger) than predicted by the models, At lower temperatures, radii tend to be smaller than what one would expect for a giant planet of solar composition.  Several authors, e.g., \citet{2011MNRAS.410.1631E}, have noted the trend that is evident in Figure~\ref{fig:Giants_AnomsvsT}, and a power-law fit to the current data suggests an ${\cal R}\propto T_{\rm eq}^{1.4\pm0.6}\,$ dependence.

Giant planets with negative ${\cal R}$ probably contain a substantial fraction of heavy elements. The majority of observed planets with $T_{\rm eq}<1000$ K have negative radius anomalies; for this population $\bar{\cal R}=-0.12$, which suggests
that the average short-period giant planet has a $M_{\rm core}\sim 30$ M$_{\oplus}$ core mass. Planets orbiting metal-rich stars are statistically associated with larger core masses \citep{2011ApJ...736L..29M}, and  there is likely an inverse relation between planetary mass, $M_{\rm P}$, and the degree of a planet's enrichment in heavy elements, $Z_{\rm p}/Z_{\star}$ \citep{2011ApJ...729L...7L}.

In order to inflate an H-He dominated planet to large ${\cal R}$, models must appeal to a cryptic source of heating at adiabatic depth, and a wide range of heating mechanisms have been proposed. Those that have received significant attention include dissipation stemming from (1) tidal orbital circularization \citep{Bodenheimer01}, (2) obliquity tides \citep{Winn05}, but see the cautions raised by \citet{2007A&A...462L...5L}, and see \citet{2007ApJ...665..754F} for full-blown arguments against obliquity tides, (3) ``kinetic heating" in which wind energy is converted into heat \citep{Guillot02}, (4) dissipation induced by thermal tides \citep{2010ApJ...714....1A}, and (5) Ohmic dissipation \citep{2010ApJ...714L.238B}.

A clue is provided by the ${\cal R}\propto T_{\rm eq}^{1.4}$ dependence. It is difficult to understand this correlation if the radius anomalies stem primarily from tidal heating, for which one would expect the major dependence to be with eccentricity, ${\cal R} = {\cal R}(e)$. Individual planets, of course, may be significantly affected by dissipation, and there is a wide observed dispersion in planetary size at all $T_{\rm eq}$.

\citet{2010ApJ...714....1A} revived the thermal tidal torque mechanism that was introduced by \cite{1969Icar...11..356G} to account for Venus' low-frequency retrograde spin. In \citet{2010ApJ...714....1A}'s picture, uneven global heating of a short-period planet produces a comparatively cold, high-density bulge that trails the star-planet line. Gravitational torque acting on the bulge spins the planet up into a state of asynchronous rotation, enabling steady-state constant-lag tidal energy dissipation to occur in the planet. \citet{2010ApJ...714....1A} suggest that this dissipation may be sufficient to inflate hot Jupiters to their observed radii. Other workers, however, e.g., \citet{2009MNRAS.395..422G}, argue that rapid isostatic re-adjustment of the planet will offset the thermal bulges for forcing frequencies (e.g., the planetary mean motion, $n= 2 \pi/P$) that are lower than the planetary dynamical frequency, which is of order the free-fall time scale, $\sqrt{\rho/G}$.

The kinetic heating mechanism, as outlined by \citet{Guillot02}, posits that a constant fraction, $\eta\sim 0.01$, of the total flux received by a planet is converted into kinetic energy by atmospheric gradients, and that this energy is dissipated at adiabatic depth within the planet. While the flow patterns on the surfaces of strongly irradiated extrasolar planets are both unknown and a matter of debate (see below), an overall $\dot{E}\propto T_{\rm eq}^{4}$ scaling is implied by the kinetic heating hypothesis, leading to ${\cal R}\propto T^{2/3}$, which is somewhat weaker than the observed ${\cal R}\propto T^{1.4}$ dependence. As was the case with the evaluation of the viability of tidal heating, this argument does not imply that kinetic heating is absent, but rather, that it is unlikely the primary cause of the radius anomalies.

\citet{2010ApJ...714L.238B} argue that ohmic dissipation inflates hot Jupiters. In their model, significant atmospheric electrical conductivity at $T_{\rm eq} \gtrsim 1200$ K is enabled by valence electron ionization of alkali metals. The presence of Na in the atmosphere of HD 209458b has been directly observed via transit spectroscopy \citep{2002ApJ...568..377C}. The Batygin-Stevenson model assumes that the atmospheric layers of the swollen planet have rotation-induced zonal jets, which are seen on all four Solar System giants, and which are a generic feature in two-and three-dimensional hydrodynamic simulations of exoplanetary surface flows \citep{2011ApJ...738...71S}. Conductive atmospheric zonal flows introduce a meridional surface current that, in the Batygin-Stevenson model, flows from the planetary poles to the equator, and then closes the circuit by returning poleward through the deep convective interior of the planet. Resistive heating within the planet, especially in the near-surface (but still adiabatic) layers, provides an internal energy source.  \citet{2011ApJ...738....1B} find that this heating is sufficient, for reasonable planetary atmospheric compositions and magnetic field values, to provide ${\cal R}$ values of order those that are observed. 

If one further assumes that the planetary Elsasser number, $\Lambda=\sigma_{i} B^{2}/2\rho\Omega$, is roughly constant across the aggregate of transiting planets, the Batygin-Stevenson theory implies an ${\cal R}\propto T_{\rm eq}^{2.7}$ dependence. This power-law index $\alpha=2.7$ is steeper than the observed value of  $\alpha\sim1.4$. A more sophisticated treatment, however, must account for the back reaction onto the wind velocity, $\bf v$, generated by the Lorentz force. Three-dimensional simulations by \citet{2013ApJ...764..103R} suggest that the Lorentz force provides a drag term to the velocity field, which in turn reduces the current and decreases the dissipation rate, driving the power-law index $\alpha$ to lower values. In summary, it appears likely that Ohmic dissipation may provide the bulk of the explanation for the large radius anomalies that characterize many short-period hot Jupiter type planets.

The realm of the gas giants has a lower mass bound of about 50 M$_\oplus$. Figure \ref{fig:MassesVsDensities} (which charts masses versus densities) suggests that for $M_{\rm P}<50$ M$_\oplus$, the overall fraction of refractory material increases steadily toward lower masses. This observed trend is in part due to selection effects -- the lowest mass planets for which both $R$ and $M$ can be measured have short periods and therefore, very high surface temperatures. As an example, Kepler-78 b \citep{2013Natur.503..381H, 2013Natur.503..377P} has $M_{\rm P} \sim1.7$ M$_{\oplus}$, $R_{\rm p}\sim1.1$ R$_{\oplus}$, and $\rho_{\rm p}\sim5.5\,{\rm g \,cm^{-3}}$, and must therefore be composed largely of metals and silicates, but its orbital period is only $P=0.35$ days, and it is so close to its star that it would  very likely have lost any low-density component to its original mass. Indeed, the low-$T_{\rm eq}$, low-$M_{\rm P}$ planets shown in Figure \ref{fig:MassesVsDensities} whose masses have been determined through analysis of transit timing variations, and which are thus less sensitive to a bias toward very short periods, have systematically lower  densities than those which have had their masses determined via Doppler velocity measurement \citep{2013DPS....4530007J}.

Bulk densities give an important clue to the overall planetary structures, but, unfortunately, there are a wide range of compositional structures involving admixtures of metals, rock, ices, and gas that can reproduce a given planetary mass and radius. See, e.g., \citet{2007ApJ...659.1661F} for a catalog of planetary structural models spanning a broad range of compositional assumptions.

The tidal Love number, $k_2$ provides an effective parameterization of the degree of central condensation exhibited by a particular density structure, $\rho(r)$, with lower values for $k_2$ mapping to higher degrees of central condensation. \citet{2009ApJ...698.1778R} introduced and outlined the practical problem of measuring $k_2$'s for individual extrasolar planets. For hot Jupiters with the shortest periods, they show that the planetary interior induces apsidal precession rates of up to a few degrees per year. 

If a short-period planet in a co-planar, hierarchical ($a_c \gg a_b$), two-planet system has tidally evolved to an eccentricity fixed point (see, e.g., \citealt{wugoldreich02} and  \citealt{mardling07}), then the outcome of this dissipative, deterministic process is that the apsidal lines of the two planets are forced to precess at the same rate.  To a good approximation, when a fixed point configuration arises, the secular gravitational perturbation from the inner planet, b, fully dictates the apsidal precession rate, $\dot{ \varpi_{c}}$, of the outer planet, c. This precession is likely too subtle to be directly observable in a radial velocity time series alone, but it is nonetheless a fully determined function of directly measurable orbital parameters, $a_b$, $a_c$, $e_b$, $e_c$, $\varpi_{b}$, $\varpi_{c}$, and $n_c$, in combination with the masses $M_{b}$ and $M_{\star}$. The precession rate, $\dot{ \varpi_{b}}$ of the inner planet, on the other hand, is dictated by a number of contributions \citep{Batyginetal09}, including secular perturbations from the outer planet, general relativity, and the rotational and tidal bulges of both the inner planet and the star. In the common case where the stellar bulges are negligible, and for an inner planet whose orbital figure, mass, and radius have all been determined via high-quality photometric transit and Doppler-based observations, $k_2$ constitutes the only significant unknown. 

As a consequence, for systems with favorable orbital configurations, one potentially has access to a mechanism that directly probes the interior structure, and which can potentially discriminating between ``super-Earth'' and ``sub-Neptune'' configurations. To date, this method has only been applied to the HAT-P-13 b-c system \citep{Batyginetal09}, where the inner planet mass is $M_{b}=0.85\pm0.038\,{\rm M_{\rm J}}$, well above the $M_{\rm P}<50$ M$_{\oplus}$ scale of interest here. \citet{Krammetal12} find $0.265 <k_2< 0.379 $  for HAT-P-13b, which requires that any central core in the planet have mass $M_{\rm core}<27$ M$_{\oplus}$. The  TESS mission \citep{Rickeretal10}, and CHEOPS mission\footnote{\url{http://cheops.unibe.ch/}} which are scheduled for launch in 2017, and the PLATO Mission \citep{2013ASPC..478..115B}, selected in February 2014 and planned for launch in 2024, may detect nearby transiting multiple-planet systems for which the orbital properties can be determined with enough precision to enable accurate measurement of $k_2$. Finally, implicit evidence for the widespread existence of dense planetary cores is provided by objects such as CoRoT 7b \citep{2009A&A...506..287L} and Kepler-10b \citep{2011ApJ...729...27B}. For a detailed overview of the types of systems for which interior structures can be probed, see \citet{2013ApJ...778..100B}.

\subsection{Observations of Exoplanetary Atmospheres and their Interpretation}

The foregoing discussion makes it clear that the radii and interior structures of the hot Jupiters continue to provide significant challenges to a comprehensive theoretical understanding. An analogous situation also holds for the atmospheric characterization of these planets.

Nevertheless, despite the extraordinary difficulties inherent in separating the signal of a faint planet from a bright star, there have been tangible successes, the majority of which have come from space-based measurements. The Spitzer Space Telescope \citep{2004ApJS..154....1W} was designed prior to the detection of transiting extrasolar planets, and exoplanetary observations were not part of the mission's scientific requirements. Serendipitously, however, Spitzer has played a key role in probing exoplanetary atmospheres. HST, MOST, CoRoT and \ik have all also made substantial contributions, as have ground-based telescopes including Keck and the VLT. The literature on the characterization of the atmospheres of short-period planets is now extensive, and space considerations preclude an exhaustive summary here. See \citet{2010ARA&A..48..631S}, \citet{2011exha.book.....P}, \citet{2012ApJ...749...93L},  \citet{2013A&ARv..21...63T}, and \citet{2014arXiv1402.1169M}  for successively recent overviews. It is important to stress that while the accumulated catalog of measurements is by now impressive, the signal-to-noise of individual measurements is often low, systematic errors may, in general, be significant, and the number of unknowns, including composition, meteorology and chemistry that characterize an atmosphere is daunting. 

Atmospheric characterization has mostly been carried out for planets that transit. Extant observations divide into (1) occultation studies, (2) transit spectroscopy of varying resolution, and (3) elaborations on full-phase spectrophotometry.

A wavelength-dependent decrease in flux can be measured during {\it occultation} (also known as secondary eclipse), when a planet passes behind a star as seen from Earth.\footnote{Some planets, such as HD~17156~b, with its highly eccentric $e=0.68$, $\omega=112^{\circ}$ orbit, undergo transit, but {\it not} occultation.} At visible wavelengths, the occulted light from the planet consists of reflected and scattered starlight, and for hotter atmospheres, emission from the planet itself. The primary conclusion is short-period, hot Jupiters are {\it dark}. For example, \citet{2008ApJ...689.1345R} used the MOST telescope to find that the geometric albedo, $A_{\rm g}$ for HD 209458 is $A_{\rm g}=0.038\pm0.045$ in the $\lambda=\,400 \rightarrow 700$ nm optical bandpass. More recently, \citet{2013ApJ...772L..16E} used the STIS instrument on HST to obtain an optical and near-UV spectrum (essentially a reflection spectrum) for HD 189733b. They measured a geometric albedo, $A_{\rm g}=0.40\pm0.12$ in the $\lambda=\,290 \rightarrow 450$ nm region, and  $A_{\rm g}<0.12$ for $\lambda=\,450\rightarrow 570$ nm (consistent with the Rowe et al. result for HD 208458b). The planet's apparent upturn in reflectivity at short wavelengths is suggestive of a cerulean visual appearance, and is attributed to starlight reflecting from clouds coupled with high-altitude sodium absorption of red light within the overlying clear air.

At near- and mid-infrared wavelengths, in the Rayleigh-Jeans tail of the star's emission spectrum, a typical hot Jupiter has a flux ratio, $F_{\rm p}/F_{\star} \gtrsim 10^{-3}$, and Spitzer has proved to be very well suited to detecting exoplanets in occultation. During the cryogenic phase of its mission, Spitzer's IRAC, IRS, and MIPS instruments were used to observe 15 exoplanets in occultation, starting with MIPS detection of HD 209458~b at 24 $\mu$m \citep{2005Natur.434..740D}, and the IRAC detection of  TrES-1 at 4.5 $\mu$m and 8 $\mu$m  \citep{2005ApJ...626..523C}. Following depletion of its liquid helium, Spitzer has continued to observe occultations with IRAC at 3.6 $\mu$m and 4.5 $\mu$m, and over thirty planets now have eclipse depths measured in both of these bands.

Photometric occultation measurements at a range of different bandpasses amount to a low-resolution dayside emission spectrum of an extrasolar planet, and a handful of planets have been measured in five or more bands, including J (1.2 $\mu$m), H (1.6 $\mu$m), and  K (2.2 $\mu$m) from the ground, e.g., \cite{2011AJ....141...30C}, in addition to the Spitzer channels. For the most favorably detected transiting planets, such as HD 189733b, Spitzer's IRS instrument was used to obtain mid-infrared dayside emission spectra that were resolved into several dozen spectral resolution elements in the 5$\,\mu$m to 14$\,\mu$m range, with each spectral element having S/N$~\sim10$, \citep[e.g.,][]{2008Natur.456..767G}. Also for HD 189733b, HST's NICMOS has been used to obtain near-IR spectra presenting 18 spectral channels in the 1.5$\,\mu$m to 2.5$\,\mu$m range, with S/N in each element also of order 10 \citep{2009ApJ...690L.114S}. These resolved spectra from IRS and NICMOS display significant variations that are strongly suggestive of molecular lines (including $\rm{H}_2$O, N${\rm H}_4$, and CO), yet fitted multi-parameter models are imperfect, suggesting a substantially incomplete picture of the radiative structure, chemistry, and dynamics of the atmosphere.

Naturally, there has been a vigorous effort to interpret the panoply of occultation results with theoretical models. For example, \citet{2008ApJ...678.1419F} suggest that strongly irradiated atmospheres (corresponding roughly to the half of the observed planets that receive the largest orbit-averaged fluxes) have thermal inversions caused by hardy molecules such as TiO or VO absorbing and re-radiating starlight at low pressures high in the atmosphere. \citet{2010ApJ...720.1569K} present evidence for an empirical correlation showing that chromospheric activity inhibits the formation of such thermal inversions, perhaps by destroying the inversion-producing molecules through an elevated flux of high-energy photons. \citep{2012ApJ...758...36M},  \citep[see  also][]{2010ApJ...725..261M}, have suggested that C/O ratios form an additional dimension by which planets can exhibit planet-to-planet variation in occultation depths. 

\citet{2013ApJ...775..137L} discuss a Bayesian retrieval approach -- see, e.g. Rodgers (2000) -- to spectral modeling which uses nonlinear optimal estimation to determine atmospheric temperatures, compositions and structures. This technique quantifies the relative information content of a given data set, and provides an estimate of the number of parameters (e.g. distinct molecular constituents) that the data can usefully be employed to determine. Spectral retrieval has been used to address a number of controversies of interpretation, for instance, in connection with the apparent presence of methane in the atmosphere of HD~189733b \citep{2014ApJ...784..133S}.

\begin{figure}[ht]
\includegraphics[height=9cm,angle=0]{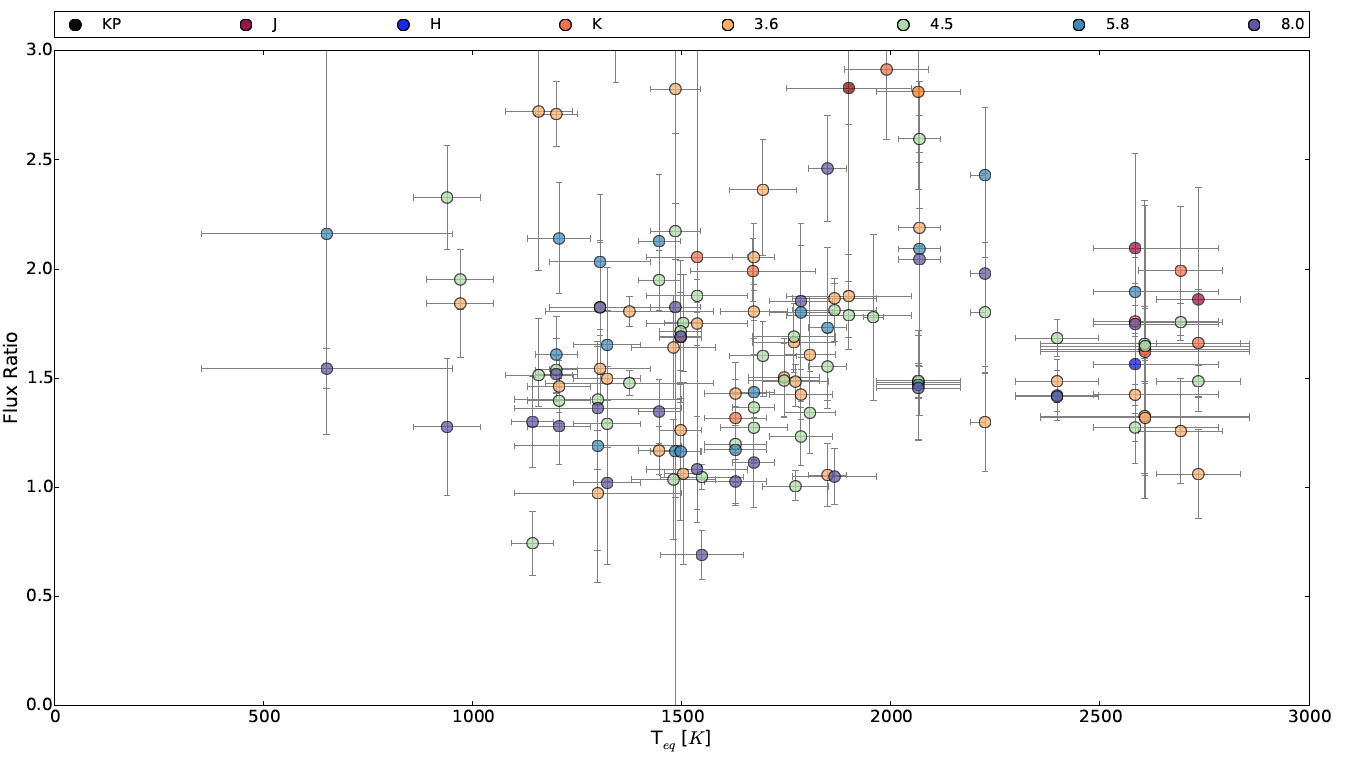}
\caption{A collection of broad-band occultation measurements. Planets are ordered on the $x$-axis by the equilibrium temperatures, $T_{\rm eq}=({R_{\star}^{1/2} T_{\star} )/({(2a)^{1/2}(1-e^{2})^{1/8}}})$, that they would have if they were $A_g=0$ blackbodies re-radiating from the full planetary surface area. Measured occultation depths are expressed as ratios of the observed eclipse depth in the band of interest relative to the expected depth in that band if the planet were a uniformly re-radiating $A_g=0$ blackbody. The color coding for each band ass are given in the legend and range from Kepler observations in the optical to 8~$\mu$m. }
\label{fig:SecondariesFigure.png}
\end{figure}

Figure \ref{fig:SecondariesFigure.png} consolidates a full range of broad-band occultation results onto a single plot. The planets are ordered according to specific orbit-averaged received flux, as measured by $T_{\rm eq}$. Occultation depths in the various infrared bands are transformed to flux ratios, $F_{\lambda}$ relative to the expectation for a zero-albedo $4\pi$ blackbody re-radiator. It is clear from the diagram that there is little or no systematic degree of similarity among the planets. The data are consistent with  random flux ratios relative to $<{F_{\lambda}}>=1.5$, with standard deviation, $\sigma_{\lambda}=0.5$. This distribution of measurements suggests a complex mix of contributing factors (stemming either from conditions on the planets, or from errors introduced in the measurement pipelines), with the attendant applicability of the central limit theorem. The $<{F_{\lambda}}>=1.5$ mean value, however, implies that on the whole, exoplanets are partially redistributing their received flux via re-radiation from the night side. Indeed, Figure \ref{fig:SecondariesFigure.png} is in general concordance with the view expressed by \citet{2014MNRAS.444.3632H} that, to the precision of the measurements, the observed emission spectra of exoplanets are generally consistent with blackbodies.

During transit, starlight flows through the atmosphere of the planet all along the day-night terminator. A stellar spectrum obtained during transit thus contains an admixture of the planet's transmission spectrum, which holds out the possibility of obtaining a detailed probe of the atmospheric column. An early application of transit spectroscopy was presented by \citet{2002ApJ...568..377C}, who used the STIS spectrograph on HST to detect sodium in the atmosphere of HD 209458b by noting that the transit depth associated with a band pass covering the sodium resonance doublet at 589.3 nm was deeper than in adjacent bands. HST has also been used to detect evidence of water vapor, methane, and carbon monoxide \citep[e.g.][]{2009ApJ...690L.114S, 2009ApJ...704.1616S} in the $1.5 \rightarrow 2.5\,\mu$m transit spectra of HD 209458b and HD 189733b. 

For planets transiting $V>8$ primaries, a variety of broad-band transmission spectra have been obtained for a variety of planets. Interpretation of these spectra is challenging, and, in parallel with the situation for occultations, firm conclusions are hampered by intrinsically low spectral resolution, low signal-to-noise, and the need to adopt a whole range of model assumptions. For more detail, the reader is advised to consult the review by \citet{2013A&ARv..21...63T}. JWST is expected to have significant utility for transit spectroscopy, and optimism exists that it will be able to usefully prove the atmospheres of potentially habitable planets transiting M-dwarf stars \citet{2009PASP..121..952D}.

There is an exciting possibility of directly detecting planetary surface winds by observing the Doppler shift of spectral absorption lines that are creating by the starlight shining through moving parcels of air at the planet's day-night terminator. \citet{2010Natur.465.1049S} used the VLT to obtain a high-resolution transit spectroscopy of HD 209458b with $R\sim100,000$ at wavelengths near $\lambda=2\mu$m. By cross-correlating on dozens of expected CO lines, these authors were able to produce a time-dependent measurement of the average velocity of the atmosphere at the planetary terminator during the course of the transit. With the orbital motion subtracted out,  \citet{2010Natur.465.1049S} measure a net blueshift of $2\pm1\,{\rm km~s^{-1}}$, and suggest that this represents a global flow from the dayside hemisphere to the nightside hemisphere at atmospheric levels in the range $0.01\,{\rm bar}<P<0.1\,{\rm bar}$. \citet{2013ApJ...762...24S} find that generic flow patterns with this large net blueshift can be created within current extrasolar planet atmospheric circulation simulations (see, e.g., \citealt{2009ApJ...699..564S}), by varying the amount of frictional drag in the atmosphere.

For planets in the ungiant regime, atmospheric measurements obtained via transit spectroscopy can potentially sort out the compositional degeneracies that prevent strong conclusions being drawn from bulk planetary density measurements (e.g., those shown in Figure \ref{fig:MassesVsDensities}). A planet with a refractory core covered by a deep solar-composition atmosphere will have a wavelength-dependent transit depth that can be potentially distinguished, using current observational sensitivity, from the signature produced by a planet with an atmosphere (and an overall structure) dominated by high molecular weight species such as water.  For example, if the atmosphere of a planet such as Gliese 1214b, with  $M_{\rm P}=6.6$ M$_{\oplus}$, $\rho_{\rm pl}=0.35~\rho_{\oplus}$, and $T_{\rm pl}\sim 500\,{\rm K}$ \citep{Charbonneauetal09}, is primarily composed of  ${\rm H}_2$, then the resulting atmospheric scale height, $H=k_{B}T/\mu g$, will be of order  $H\sim200\,{\rm km}$. Given that Gliese 1214b has an M4.5 V primary, this implies that  $H/R_{\star}\sim0.0014$. For an $H$ to $R_{\star}$ ratio of this magnitude, ground-based measurements are presently capable of distinguishing between a clear $H_2$-dominated solar-composition atmosphere, which will show measurable differences in planetary radius as a function of wavelength, $\Delta(R_{\rm p}/R_{\star})\sim0.005$ \citep{deMooijetal12}, and a high-molecular weight atmosphere with a small scale height and a transmission spectrum that presents much smaller variations.

Measurements of $\Delta(R_{\rm p}/R_{\star})$ taken at different bandpasses during transits by Gliese 1214b have indicated that the transmission spectrum is fairly flat between $\lambda=0.6~\mu$m and $\lambda=5~\mu$m \citep{Beanetal10,Desertetal11,Crolletal11, Beanetal11, Bertaetal12, deMooijetal12}. As of 2014, number of these measurements are in conflict, however, and while a clear solar-composition atmosphere is inconsistent with the data, Gliese 1214b could be presenting either a hazy low-metallicity H-He dominated atmosphere, or a low-scale height atmosphere composed of species (such as ${\rm H_{2}O}$) of high molecular weight.  \citet{2014Natur.505...69K} report intensive  HST observations of Gliese 1214b (15 transits over 60 telescope orbits) that are sensitive enough to reveal absorption features from constituents such as ${\rm H_{2}O}$, ${\rm CO_{2}}$, and ${\rm NH_{4}} $ in a high molecular weight atmosphere. A flat transmission spectrum between 1.1 $\mu$m and 1.7 $\mu$m was observed, suggesting that Gliese 1214b's atmosphere {\it must} contain clouds or high-altitude hazes. In addition, other planets with $P<100\,{\rm d}$ and $M<30$ M$_{\oplus}$ are nearly certain to be discovered in transit across bright low-mass primaries. For example, \citet{Bonfilsetal12} have announced the detection of Gliese 3470b, which has $M_{\rm P}=14$ M$_{\oplus}$, $R_{\rm p}=4.2 \,$ R$_{\oplus}$, and $T_{\rm pl}\sim 800\,{\rm K}$, and which orbits an $R_{\star}\sim0.5\,$ R$_{\odot}$ M1.5 V primary. Spectrographic observations of of Gliese 3470b by \citet{Crossfieldetal13}  using the Keck telescope are consistent with a flat spectrum between 2.09 $\,\mu$m and 2.36$\,\mu$m, which rule out a cloud-free atmosphere in chemical equilibrium, and which are again suggestive of high-altitude haze or clouds. The prospects for improved characterization of planets in the ungiant category are especially bright, given that NASA's forthcoming TESS Mission will locate effectively all of the transiting short period planets with radii $R_{\rm p} \gtrsim 1.5$ R$_\oplus$ orbiting stars with $V<12$.

\begin{figure}[ht]
\centering
\includegraphics[height=10.5cm,angle=0]{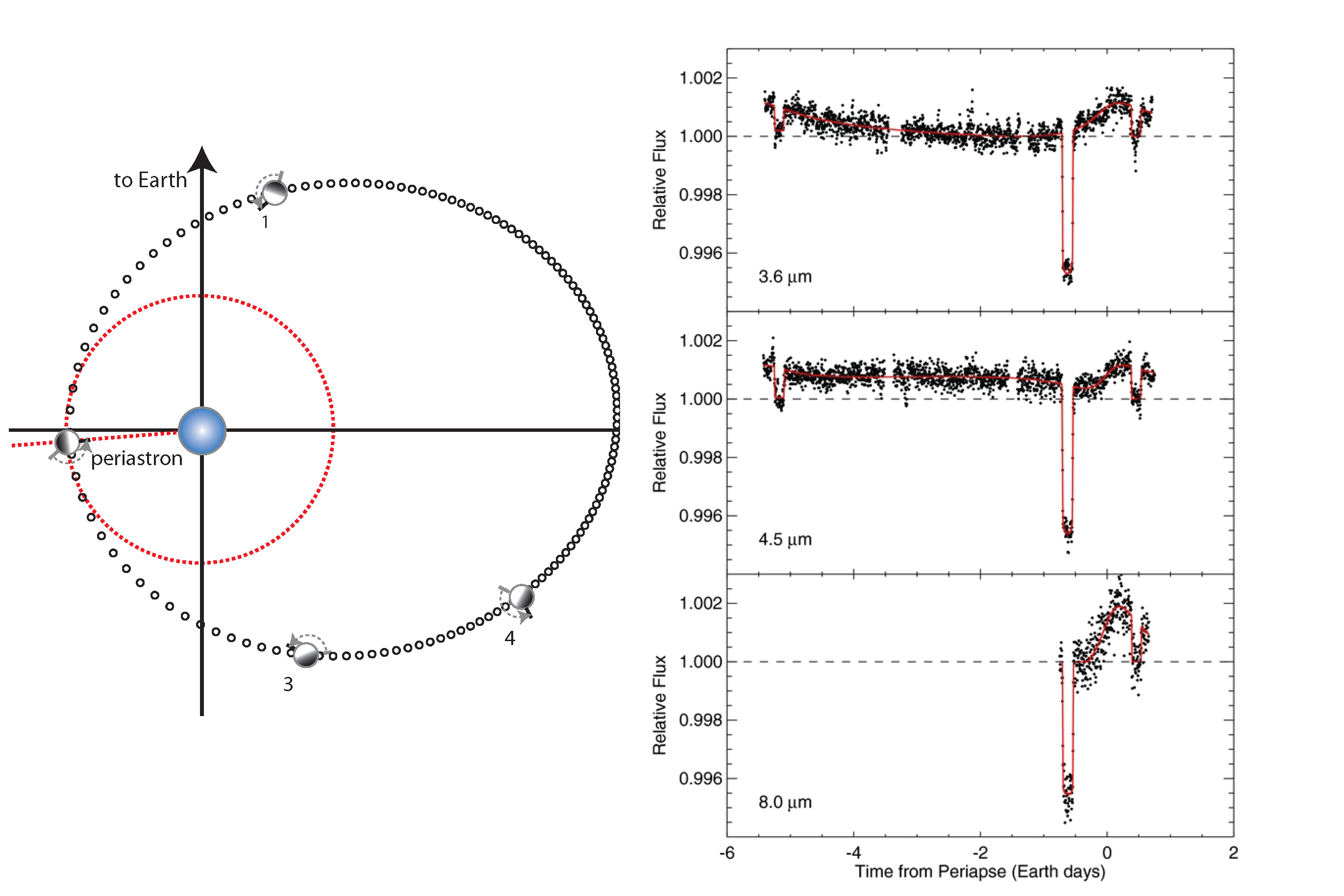}
\caption{Photometric IR phase curves for the eccentric planet HAT-P-2b,  with orbital period, $P$=5.633d, eccentricity, $e$=0.511, and $M=9M_{\rm Jup}$, adapted from \citet{2013ApJ...766...95L}. The top panel shows the geometry of the HAT-P-2b orbit, with periastron passage occurring 14 hours after the primary transit. The lower panels display the light curves for the 3.6 (top), 4.5$\,\mu$m (middle), and 8.0 $\,\mu$m photometry (filled circles). Optimized phase, transit, and secondary eclipse fits are over-plotted as a red line, see \citet{2013ApJ...766...95L} for details. The dashed line represents the stellar flux level.}
\label{fig:hatphase}
\end{figure}

Transit and occultation measurements act as effective bookends to {\it full-phase} photometric measurements, in which planetary emission is monitored through the course of an orbit. The most frequently cited example of this type of observation was published by \citet{2007Natur.447..183K}, who presented a high-quality 8$\,\mu$m light curve for HD 189733b that spans the 29-hour interval from just before the planet's primary transit until just after its secondary eclipse. Adopting the assumption that the planet is in spin-synchronous rotation with its orbit, \citet{2007Natur.447..183K} derived a global temperature distribution for the planetary photosphere as a function of longitude, and found that the hottest region of the planet is  shifted east of the substellar point  (see \citet{2008ApJ...678L.129C} for the details of the mapping technique). The longitudinal shift of the planet's temperature maximum is attributed to super-rotating ${\bf v}\sim1\,{\rm km\,s^{-1}}$ equatorial winds, which are a robust feature that emerge from a wide variety of hydrodynamical models of the surface flows on tidally locked close-in extrasolar planets, \citep[see, e.g.][]{2009ApJ...699..564S, 2011ApJ...738...71S}. In the past several years, a number of additional infrared phase curves in assorted bandpasses have been produced for a variety of strongly irradiated planets. Examples include \citet{2009ApJ...690..822K}'s follow-on partial-orbit MIPS 24$\,\mu$m and \citet{2012ApJ...754...22K}'s  full-orbit 3.6$\,\mu$m and 3.5$\,\mu$m IRAC phase curves for HD 189733   \citep[which corroborate the basic picture inferred by][]{2007Natur.447..183K}, and a full-orbit optical phase curve from the Kepler satellite for HAT-P-7b \citep{2009Sci...325..709B} that shows evidence of both dayside thermal emission and reflected light. Full or partial-orbit Near-IR phase curves have also been published for HD 149026b \citep{2009ApJ...703..769K}, WASP-18b \citep{2013MNRAS.428.2645M}, and WASP-12b \citep{2012ApJ...747...82C}, all of which are hot, short-period planets on nearly circular orbits. The phase curves for these systems suggest that the selection of planets surveyed have a large range of efficiency in their ability to equilibrate day-night temperature distributions.

A typical hot Jupiter with $T_{\rm eq}\sim1500$K has its blackbody emission peak at $\lambda\sim1.9\mu{\rm m}$, which is blueward of the Spitzer bandpasses. It is therefore useful to obtain phase curves near the region where a given planet disgorges the majority of its energy. \citet{2014Sci...346..838S} used the HST WFC3 instrument to obtain full-phase 15-channel spectrophotometry in the  $1.1\mu{\rm m} \rightarrow 1.7\mu{\rm m}$ region for three consecutive full orbits of WASP~43b. This planet is a representative hot Jupiter, albeit with a cool ($T_{\rm eff}=4500$, K7) primary, and a short $P=19.5$h orbit. Stevenson et al.'s analysis indicates that the phase-resolved spectrum shows little contribution from molecular absorbers, with the only significant influence coming from $\rm{H_{2}O}$. In agreement with many of the other extant full-phase observations, the light curve indicates that the planet's hottest location is offset eastward from the substellar point, and that there is poor dayside -- nightside heat circulation.

Figure \ref{fig:hatphase} shows phase curves in several bandpasses for the eccentric planet HAT-P-2b, as presented by \citet{2013ApJ...766...95L}. Planets with eccentric orbits cannot rotate synchronously, and hence their atmospheres are virtually certain to display strongly time-dependent meteorology. Phase-curve observations of such planets can potentially determine atmospheric radiative time constants, tidal luminosities, and pseudo-synchronous rotation rates, as well as the global manifestation of time-dependent storms \citep{2009Natur.457..562L}.

\section{Planet Formation}

The vast majority of work on planetary formation has been done to explain the properties of our Solar System, for which we have the most detailed data.  We begin with a synopsis of these models, and then discuss some recent models of the formation and migration of exoplanets that travel on short-period orbits.

\subsection{An Overview  Informed by the Solar System}

Models of Solar System formation are based upon data from our own Solar System as well as observations of the circumstellar environments of young stars.  The nearly planar and almost circular orbits of the planets in our Solar 
System argue strongly for
planetary formation within flattened circumstellar disks.  When a molecular cloud core 
undergoes gravitational
collapse, an occurrence triggered by loss of magnetic field and turbulent velocity support, its spin angular momentum leads to the formation of 
pressure-supported protostars
surrounded by rotationally-supported disks, e.g., \citet{1984ApJ...286..529T}. Molecular cloud cores are sub-condensations of mass, $M\gtrsim $~M$_{\odot}$, characteristic size $r\sim 0.25$~pc, and rotation rate typically $\omega \sim 10^{-13}\,{\rm Hz}$ that are found within giant molecular cloud complexes in the galactic disk that typically contain $10^{4}$ to $10^{5} $~M$_{\odot}$. 

Circumstellar disks are commonly observed in association with newly formed stars \citep{2009AIPC.1158....3M}, and  are analogous to 
the primordial solar nebula that was initially conceived by Kant and Laplace to explain the 
observed properties of
our Solar System; for a more detailed historical background, see \citet{1993QJRAS..34....1W}. Planets form within such disks, and play a major role in the later stages of disk evolution.  Terrestrial planets grow via pairwise accretion of solid bodies.  In the generally favored core-accretion paradigm, giant planets begin their growth as do terrestrial planets, but they become massive enough to accumulate substantial amounts of gas before the protoplanetary disk dissipates.  Under some circumstances, as described for example, by \citet{2001ApJ...553..174G}, giant planets may form directly from the protostellar disk via gravitational instability
\citep[see, e.g.][]{2007prpl.conf..607D, 2009ApJ...695L..53B}, especially at large ($d\gtrsim40$AU) from the parent star.

Within the Solar System, bodies up to the mass of Earth consist almost entirely of condensable
(under reasonable protoplanetary disk conditions) material, and the fraction of highly
volatile gas increases with mass above that of Earth. The most massive planets contain a considerable fraction of light gases.  About $90\%$ of Jupiter's mass is H and He, and these two light elements make up $\sim 75\%$
of Saturn.  In both cases, these planets contain $\sim 30~{\rm M}_{\oplus}$ in elements other than hydrogen and helium. The large amounts of H and He contained in Jupiter and Saturn imply that these planets must
have formed within $\sim 10^7$ years of the collapse of the Solar System's natal cloud, before
the gas in the protoplanetary disk was swept away. The two largest planets in our Solar System are generally referred to as {\it gas
giants}, even though these elements aren't gases at the high pressures that most of the
material in Jupiter and Saturn is subjected to. Analogously, Uranus and Neptune are frequently
referred to as {\it ice giants}, even though the astrophysical ``ices'' such as H$_2$O, CH$_4$,
H$_2$S and NH$_3$ that models suggest make up the majority of their mass \citep{1995netr.conf..109H} are
in fluid rather than solid form. Note that whereas H and He {\it must} make up the bulk of
Jupiter and Saturn because no other elements can have such low densities at plausible
temperatures, it is possible that Uranus and Neptune are primarily composed of a mixture of
``rock'' and H/He, although the high observed CH$_4$/H$_2$ ratio in their atmospheres (30-40$\times$ the solar value) makes this somewhat unlikely . This ambiguity in basic structure also holds for the majority of the ungiant exoplanets whose mass and radius have both been measured.

Lithium and heavier elements constitute $< 2\%$ of the mass of a solar composition mixture.
The {\it atmospheric} abundances of volatile gases heavier than helium (excluding neon, which is thought to be substantially depleted through gravitationally-induced settling) are $\sim 3 \times$
solar in Jupiter \citep{2003NewAR..47....1Y}, a bit more enriched in Saturn, and substantially more for
Uranus and Neptune. The {\it bulk} enhancements in heavy elements relative to the solar value
are roughly 5, 15, and 300 times for Jupiter, Saturn and Uranus/Neptune, respectively. Thus,
all four giant planets accreted solid material substantially more effectively than gas from
the surrounding nebula. Moreover, the total mass in heavy elements varies by only a factor of
a few between the four planets, while the mass of H and He varies by about two orders of
magnitude between Jupiter and Uranus/Neptune.

There is a vast observational literature that details the presence, the properties, and the evolution of disks
of Solar System dimensions around pre-main sequence stars \citep{2011ARA&A..49...67W}. The existence of disks on scales of a 
few tens of
astronomical units is inferred from the power-law spectral energy 
distribution in the infrared
over more than two orders of magnitude in wavelength 
\citep{chianggoldreich97, 2001ApJ...553L.153H, 2004ApJS..154..367M, 2010ApJ...723.1241A}. Observations of
infrared excesses in the spectra of young stars suggest that the 
lifetimes of protoplanetary
disks span the range of $10^6$ -- $10^7$ years 
\citep{1993prpl.conf..837S, 2002ApJ...571..378A, 2012arXiv1208.4689E}. Much lower mass second-generation debris disks composed of dust eroded from larger solid objects that accreted within protoplanetary disks are observed around main-sequence stars, with younger stars typically (but far from always) possessing more massive debris disks \citep{2004ApJS..154..422M, 2012AJ....144...45K}.  Many debris disks appear to have inner holes, and lower mass stars typically retain their debris disks for more time \citep{2008ARA&A..46..339W}. 

Although to a first approximation the gas in the disk 
is centrifugally supported in balance with the star's gravity, negative radial 
pressure gradients provide a small outwardly-directed force that acts to 
reduce the effective gravity, so the gas rotates about the star at slightly less than the 
keplerian velocity.  Small solid bodies (dust grains) rotate with the gas.  
Large solid bodies orbit at the keplerian velocity, and medium-sized 
particles move at a rate intermediate between the gas velocity and the 
keplerian velocity; thus medium-sized and larger bodies are subjected to a headwind from the 
gas \citep{1976PThPh..56.1756A}. This headwind removes angular momentum from 
the particles, causing them to spiral inwards towards the star.  This inward 
drift can be very rapid, especially for particles whose coupling time to the 
gas is similar to their orbital period; smaller particles drift less rapidly 
because the headwind they face is not as strong, whereas large particles 
drift less because they have a greater mass to surface area ratio.  Orbital 
decay times for meter-sized particles at 1 AU from the Sun have been 
estimated to be only $\sim$100 years \citep{1977MNRAS.180...57W}.  The large radial 
velocities of bodies in this size range relative to both larger and smaller 
particles implies frequent collisions, so most solid bodies may pass through 
the critical size range quickly without substantial radial drift.  However, it 
is also possible that a large amount of solid planetary material is lost from 
the disk in this manner.   Solid bodies larger than $\sim$1 kilometer in size face a headwind only slightly 
faster than meter-size objects, and because of their much greater 
mass/surface area ratio they suffer far less orbital decay from interactions 
with the gas in their path.  Thus,  kilometer-sized or larger planetesimals 
appear to be reasonably safe from loss (unless they are ground down to 
smaller sizes via disruptive collisions) until some of these planetesimals 
grow into planetary-sized bodies. 

The initial  composition of the protoplanetary disk is 
assumed to still be essentially identical to that of the protosun.  Dust within a protoplanetary disk initially agglomerates via sticking/local
electromagnetic forces that produce inelastic collisions. Particles gradually settle towards the disk 
mid-plane as they grow.

The conceptual bottleneck created by the short residence times of meter-sized particles in protoplanetary disks is perhaps the least well 
understood phase of solid planet formation. It is not yet clear how agglomeration from cm-sized
pebbles to km-sized or larger bodies that are referred to as planetesimals occurs. Collective gravitational
instabilities \citep{safronov69, goldreichward73} might be important, although turbulence may
prevent protoplanetary dust layers from becoming thin enough to be gravitationally unstable
\citep{weidenschillingcuzzi93}. The interactions between small particles and gas in a rotating disk involve complicated fluid-particle interactions, and planetesimal formation is a very
active research area. Recent work suggests that the answer may lie in the joint action of aerodynamic and gravitational instabilities to agglomerate particles -- see the review of \citet{chiangyoudin10}. Some models suggest that pebble-sized particles accumulated into planetesimals more than 100 km in size \citep{2012A&A...537A.125J}.

 The primary perturbations on the keplerian orbits of kilometer-sized and 
larger bodies in protoplanetary disks are pairwise gravitational interactions 
and physical collisions \citep{safronov69}.  These interactions lead to 
accretion (and in some cases erosion and fragmentation) of planetesimals.  
Gravitational encounters are able to stir planetesimal random velocities up 
to the escape speed from the largest common planetesimals in the swarm 
\citep{safronov69}.  The most massive planets have the largest 
gravitationally-enhanced collision cross-sections, and accrete almost 
everything with which they collide.  If the random velocities of most 
planetesimals remain much smaller than the escape speed from the largest 
bodies, then these ``planetary embryos" grow extremely rapidly 
\citep{1992Icar..100..440G}.  The size distribution of solid bodies 
becomes quite skewed, with a few large bodies growing much faster than 
the rest of the swarm in a process referred to as runaway accretion 
\citep{1978Icar...35....1G, 1989Icar...77..330W}.   If embryos dominate the stirring, then $\dot{M} \propto M^{2/3}$; in this circumstance, if individual embryos control the velocity in their own zones, larger embryos take longer to double in mass than do smaller ones,
although embryos of all masses continue their runaway growth relative to
surrounding planetesimals; this phase of rapid accretion of planetary
embryos is known as {\it oligarchic growth} \citep{KokuboIda98}.  Eventually, planetary 
embryos accrete most of the (slowly moving) solids within their 
gravitational reach, and the runaway/oligarchic growth phase ends. 

The self-limiting nature of runaway and oligarchic growth implies that massive
planetary embryos form at regular intervals in semi-major axis. The
agglomeration of these embryos into a small number of widely spaced
terrestrial planets necessarily requires a stage characterized by large
orbital eccentricities, significant radial mixing, and giant impacts. At the
end of the rapid-growth phase, most of the original mass is contained in the
large bodies, so their random velocities are no longer strongly damped by
energy equipartition with the smaller planetesimals. Mutual gravitational
scattering can pump up the relative velocities of the planetary
embryos to values comparable to the surface escape
velocity of the largest embryos, which is sufficient to ensure their
mutual accumulation into planets. The large velocities imply small
collision cross-sections and hence long accretion times.

Once the planetary embryos have perturbed one another into crossing
orbits, their subsequent orbital evolution is governed by close
gravitational encounters and violent, highly inelastic collisions. Terrestrial planets continue to grow by pairwise accretion of solid bodies until the spacing
of planetary orbits becomes sufficient for the configuration to be stable to gravitational
interactions among the planets for the lifetime of the system
\citep{safronov69, lissauer95, chambers01, laskar00, hansenmurray12}. 
This process has been studied using $N$-body integrations\index{N-body integrations}
of planetary embryo orbits. Many of the simulations of this type have endeavored to reproduce our
Solar System; they generally begin with about 2 M$_{\oplus}$ of material spread through
the terrestrial planet zone, typically divided (not necessarily equally) among hundreds or thousands of bodies. 
These simulations generally display a high efficiency of converting seed embryos
into terrestrial planets, and the end result is generally the formation of 2~--~5 terrestrial
planets on a timescale of about $10^8$ years.  Some of these
systems look startlingly similar to our inner Solar System, with the caveat that the emergent  planets often travel on more eccentric orbits that those of the terrestrial planets, although the simulations that begin with the largest numbers of bodies yield eccentricities more similar to those in our Solar System \citep{2006Icar..184...39O}. It is possible
that the Solar System is by chance near the quiescent end of the
distribution of terrestrial planets, but more likely, a damping process of some type plays a role. Processes such as
fragmentation and gravitational interactions with a remaining population
of small debris, and interactions with a remnant gas disk \citep{ogiharaetal07} have been invoked to lower the characteristic eccentricities
and inclinations of the ensemble of terrestrial planets.

An important result of these $N$-body simulations is that planetary
embryo orbits execute a random walk in semi-major axis as a consequence
of successive close encounters. The resulting widespread mixing of
material throughout the terrestrial planet region diminishes any chemical
gradients that may have existed when planetesimals
formed, although some correlations between the final heliocentric
distance of a planet and the region where most of its constituents
originated are preserved in the simulations. Nonetheless, these dynamical
studies imply that Mercury's high iron abundance is unlikely to
have arisen from chemical fractionation in the solar nebula.

The mutual accumulation of numerous planetary embryos into a small
number of planets must have entailed many collisions between
protoplanets of comparable size. Mercury's original
silicate mantle was probably partially stripped off by one or more of such giant
impacts, leaving behind an iron-rich core. Accretion simulations also lend
support to the giant impact hypothesis for the origin of the Earth's
Moon; during the final stage of
accumulation, an Earth-size planet is typically found to collide with
several objects as large as the Moon and frequently one body as massive
as Mars \citep{2001Icar..152..205C}.

Models for the formation of gas giant planets were reviewed by \citet{2000prpl.conf.1081W} and  \citet{2013arXiv1311.1142H}.
Star-like direct quasi-spherical collapse is not considered viable, both because of the
observed brown dwarf desert mentioned above and theoretical arguments against the formation of
Jupiter-mass objects via fragmentation \citep{Bodenheimeretal00}. The theory of giant planet
formation that is favored by most researchers is the {\it core-nucleated accretion model}, in which
the planet's initial growth resembles that of a terrestrial planet, but eventually becomes
sufficiently massive (several M$_{\oplus}$) that it is able to accumulate substantial amounts
of gas from the surrounding protoplanetary disk. The only other hypothesis receiving
significant attention is the {\it gas instability model}, in which the giant planet forms
directly from the contraction of a clump that was produced via a gravitational instability in
the protoplanetary disk \citep{2007prpl.conf..607D}.  Formation of giant planets via gas instability has never been demonstrated for realistic disk
conditions. Moreover, this model has difficulty explaining the super-solar abundances of heavy
elements in Jupiter and Saturn, and it does not explain the origin of planets like Uranus and
Neptune. Nonetheless, it is possible that some giant planets within the currently observed census did form via disk instability.

The core-nucleated accretion model \citep{1980PThPh..64..544M, pollacketal96}, has been reviewed by \citet{2007prpl.conf..591L}, with further details given in \citet{2011exop.book..319D}. The model relies on a combination of planetesimal accretion and gravitational
accumulation of gas. In this theory, the core of the giant planet forms first by accretion of
planetesimals, while only a small amount of gas is accreted.  Core accretion rates depend upon the surface mass density of solids in the disk and physical assumptions regarding gas drag, etc.~\citep{Lissauer87, Inaba03}. The escape velocity from a
planetary embryo with $M > 0.1$ M$_\oplus$ is larger than the sound speed in the gaseous
protoplanetary disk. Such a growing planetary core first attains a quasi-static atmosphere
that undergoes Kelvin-Helmholtz contraction as the energy released by the accretion of planetesimals and gas
 is radiated away at the photosphere. 

During the runaway planetesimal accretion epoch, the protoplanet's mass increases rapidly. The
internal temperature and thermal pressure increase as well, preventing substantial amounts of
nebular gas from falling onto the protoplanet. When the planetesimal accretion rate decreases,
gas falls onto the protoplanet more rapidly. The protoplanet accumulates gas at a gradually
increasing rate until its gas mass is comparable to its heavy element mass. The key factor
limiting gas accumulation during this phase of growth is the protoplanet's ability to radiate
away energy and contract  \citep{pollacketal96, 2005Icar..179..415H}. The rate of gas accretion then accelerates more rapidly,
and a gas runaway occurs. The gas runaway continues as long as there is gas in the vicinity of
the protoplanet's orbit. The protoplanet may cut off its own supply of gas by gravitationally
clearing a gap within the disk \citep{linpapa79}. \citet{lissaueretal09} used  a three-dimensional adaptive mesh refinement code \citep{2002AA...385..647D, 2003ApJ...586..540D} to follow the flow of gas onto an accreting giant planet, allowing the determination of final planetary mass as a function of the time-varying properties (density,
temperature, viscosity, longevity, etc.) of the surrounding disk. Such a self-regulated growth
limit provides a possible explanation to the observed mass distribution of extrasolar giant
planets. 

Core-nucleated accretion also offers an explanation for the formation of low-density ungiant exoplanets.  These bodies could be cores that grew massive enough to accrete some H$_2$ and He from their planetary disks, but were unable to accrete enough light gases to become giants \citep{2011ApJ...738...59R}.

\subsection{How did short-period planets arrive at their observed locations?}

Quite aside from the zeroth-order questions related to their atmospheric properties and internal structures, the origins of the short-period planets are not fully explained. It is clear from Figure 2 that the hot Jupiters represent a population distinct  from the short-period ungiants, and so it is reasonable to consider their formation scenarios  separately.

It is generally assumed that hot Jupiters were formed through the core accretion process, with assembly occurring at distances  at least a few AU from their host stars, especially water ice, were available. As discussed above, in this standard core-accretion scenario for giant planet formation, a protoplanetary core composed largely of ices and rock reaches a mass of $\sim10$ M$_{\oplus}$, at which point it begins to rapidly accrete gas from the surrounding nebular disk. Thereafter, as a result of dynamical or hydrodynamical processes, the planet is delivered into a short period orbit. It is worth noting, however, that while disfavored, {\it in situ} formation for hot Jupiters has not been strictly ruled out. Detailed calculations by \citet{Bodenheimeretal00}, for example, suggest that hot Jupiters can form in place under certain conditions.

The ``classical'' formation scenario for hot Jupiters was outlined by \citet{LBR06}, shortly after the discovery of 51 Peg b. In this picture, the planet grows to near its original mass at a distance  several AU from the star. A planet with 51 Peg b's mass is large enough to maintain a well-cleared annular gap in the disk that brackets the planet's radial location. Once formed, the planet spirals inward via the process of Type II migration \citep{1986ApJ...309..846L} on the parent protoplanetary disk's viscous time scale, $\tau_{\nu}=r_{d}^{1/2}/\nu \sim 10^{6}\,{\rm yr}$, where $\nu$ is the effective kinematic viscosity, and $r_{d}$ is the characteristic radial extent of the disk. The final location is often assumed to be controlled either by disk truncation at an inner magnetospheric cavity, or via tidal interaction with the young, rapidly spinning parent star. This basic picture has been elaborated substantially during the past two decades. For overviews, see, e.g. \citet{2011exop.book..347L} and \citet{2012ARA&A..50..211K}.

The disk migration scenario avoids the serious problems associated with {\it in situ} formation, but is not itself without problems. One difficulty lies with so-called Type I migration \citet{1997Icar..126..261W}, which is though to affect planets that have insufficient mass to clear and maintain a gap in the protostellar disk. The timescale for Type I migration $\tau_{\rm I}\sim (M_{\rm P}/$M$_{\oplus}) 10^{5}\,$yr, is substantially shorter than the timescale that appears to be required to assemble a core large enough to initiate rapid gas accretion \citep{2010Icar..209..616M}. The mismatch hints at the possibility that a fundamental process is either missing or is not fully understood. A further problem arises from observed statistics of systems with massive, short-period planets. Hot Jupiters appear almost exclusively in single-planet configurations \citep{2012PNAS..109.7982S} (with the $\upsilon$ Andromedae planets providing a notable exception). 

Purely dynamical mechanisms have received increased attention in connection with the production of hot Jupiters. An early and influential study was carried out by \citep{1996Sci...274..954R}, who attributed hot Jupiters to the outcome of catastrophic planet-planet scattering. In their picture, extant short period planets are survivors of strong gravitational encounters, in which one planet was either ejected or placed in a long-period orbit, and the other is left in a high eccentric orbit with modest semi-major axis. Given a sufficiently close periastron distance, a high-eccentricity planet will subsequently undergo orbital circularization.

The process of Lidov-Kozai migration with tidal friction has grown rapidly in popularity. In the Lidov-Kozai scenario, see e.g. \citep{2001ApJ...562.1012E, WuMurray03, FabTrem07}, a planet destined to become a hot Jupiter forms in a low-eccentricity orbit at a distance of several AU from a parent star of mass $M_{\star}$. In addition, the planetary orbit is substantially inclined at an initial angle $i_{0}$ to the (for analytic purposes) circular orbit of a distant binary companion of mass ${M_b}$. If the initial inclination angle, $i_{0}$, is larger (or better, substantially larger) than the Lidov-Kozai critical angle, $i_0>i_{\rm crit} = \arccos[(3/5)^{1/2}]=39.2^{\circ}$, the planet is compelled to undergo {\it Lidov-Kozai cycles} in which the Jacobi energy, $E_{J}=(1-e_{\rm P}^2)^{1/2}\cos i_{\rm P}$ is conserved, and in which the planet experiences periodic cycling between successive states of high orbital eccentricity and high inclination to the binary plane. 

The associated timescale for these cycles is $\tau_{\rm LK}=({2 P_{b}^{2}}/{(3\pi P_{\rm P})}) ((M_{\star}+M_{\rm P}+M_{b})/M_b)$.
For large $i_0$, the planet's periastron distance during the high-eccentricity phases of the cycle can be small enough so that dynamical tides acting on the planet efficiently convert orbital energy into heat, thereby shrinking the semi-major axis, $a_{\rm P}$, of the planetary orbit. The Lidov-Kozai cycles are eventually destroyed (or severely modified) when the 
characteristic time scale for general relativistic precession, dictated by 
$\dot{\omega}_{\rm GR}=3GM_{\star}n_{\rm P}/({a_{\rm P}}c^2(1-e_{\rm P}^2))\,$
 is substantially less than the Lidov-Kozai cycling timescale, i.e., $2\pi/\dot{\omega}_{\rm GR}<\tau_{\rm LK}$. At this point the planet can transition to a long-running phase of secular evolution that is characterized by steady orbital decay and circularization, with the end product being a hot Jupiter. 

The Lidov-Kozai process is likely to deliver planets into orbits in which the planetary orbital angular momentum and the stellar spin angular momentum are initially largely uncorrelated. A slew of measurements of the sky-projected angle, $\lambda$, between the stellar rotation and planetary orbital angular momentum vectors using the  Rossiter-McLaughlin spectroscopic technique \citep{1924ApJ....60...15R, 1924ApJ....60...22M} indicate that a substantial fraction of hot jupiter orbits are substantially misaligned with the equators of their parent stars \citep{2014arXiv1410.4199W}, yielding evidence in favor of the Lidov-Kozai mechanism as a formation channel for a subset of the observed hot Jupiter population.

\section{Questions for the Next Ten Years and Beyond}

There are few precedents in the history of science in which a discipline moves so rapidly from shaky disrepute through a golden age of discovery and into a mature field of inquiry. In less than two decades, the study of extrasolar planets has accomplished all of this -- answering questions that were posed at the dawn of the scientific era, while affording tantalizing glimpses of revelations to come. This rapid progress  partially obscures the fact that the extrasolar planets are {\it fundamentally alien}. Virtually none of their properties, either statistical or physical, aside from the scarcity of planets with $30$ M$_{\oplus}<M_{\rm P}<100$ M$_{\oplus}$, were predicted or anticipated (Lissauer 1995, 2006), and theorists struggle to understand the zeroth-order features of the planetary distribution. At the current moment, it is thus useful to conclude by listing some provocative unsolved questions.

\begin{itemize}

\item
{\it What are the bulk compositions of the ungiants?} If one considers logarithms of mass, the most prominent mass gaps among the bodies populating the Solar System are those between Jupiter and the Sun, with $\log_{10}(M_{\odot}/M_{\rm J})=3.02$, and between Earth and Uranus, with $\log_{\rm 10}(M_{\uranus}/{M_{\oplus}})=1.16$. While it has long been possible to explore the properties of objects in the Sun-Jupiter gap through observations of low mass stars and brown dwarfs \citep{2000ARA&A..38..337C}, the mere presence of extrasolar planets with $M_{\oplus}<M_{\rm P}<M_{\uranus}$ orbiting main sequence stars has been known only since the discovery of Gliese 876 d, with $M_p=7.5\,$ M$_{\oplus}$ 
\citep{2005ApJ...634..625R}, 
and the first density estimate became available only with the unveiling of Gliese 1214b, which has $M_{\rm P}=6.3$ M$_{\oplus}$, $R_{\rm p}=2.7$ R$_{\oplus}$, and $\rho_{\rm p}=1.7\,{\rm g\, cm}^{-3}$ \citet{2009Natur.462..891C}. Structural models, as calculated by, e.g., Rogers \& Seager (2010) or by Fortney et al. (2011), can easily reproduce Gliese 1214b's mass and radius with configurations drawn from three board classes of models, (i) rock and iron planets with massive H/He or ${\rm H_{2}}$ atmospheres, (ii) rock-ice cores surrounded by H-He envelopes of solar nebular composition, and (iii) so-called ``water worlds'' composed primarily of H$_2$O. 

As we have described, attempts to characterize the atmosphere of Gliese 1214b have been met with ambiguity. It is likely that only when high-quality transmission spectra and interior mass distribution measurements are obtained that a clear understanding of the true range of compositions and structures of the ungiants will be obtained.

\item
{\it What (if any) is the observable role of magnetohydrodynamics?} Magnetic fields are important in nearly every branch of astrophysics. Earth and the four larger planets in our Solar System all harbor intrinsic magnetic fields that are generated by dynamo action in the planetary interiors. It is therefore natural to expect that many, perhaps most, short-period extrasolar planets with $M_{\rm P}>$ several M$_{\oplus}$  also have magnetic fields, and to first approximation, these fields can be estimated by assuming that the Elsasser number, $\Lambda = {\sigma_i {B}^{2}/{2 \rho \Omega}}$, is roughly constant across the aggregate of transiting planets ranging from Earth-like to Jupiter-like worlds.  If we further assume synchronous rotation (which should hold for hot jupiters on circular orbits), then ${1/{\Omega}} \propto a^{3/2}$. The upper atmospheres of planets with $P\lesssim5\,{\rm d}$ are hot enough ($T\gtrsim1200\,{\rm K}$) that alkali metals (e.g., Na, K) have ionization fractions that are sufficient to couple the planetary magnetic field to the atmospheric hydrodynamic flow (see, e.g., \citealt{1992pavi.book.....S}). Studies, such as those by \citet{2010ApJ...714L.238B}, suggest that Ohmic dissipation is responsible for heating planetary interiors, thereby explaining much of the observed radius anomaly trend. Do MHD-mediated influences have a zeroth-order effect on exoplanet atmospheres, and can some of the bewildering variation in atmospheric emission be understood as arising from magnetohydrodynamic phenomena?

\item
{\it Where in the protostellar disks did the observed planets form?} Our Solar System has long been thought to be the product of planet formation that occurred more or less {\it in situ}, via core accretion for the giant planets and via planetesimal accretion for the terrestrial planets. The detection of 51 Peg b, along with other early extrasolar planet discoveries, led to the suggestion, which soon became a paradigm, that long-distance planetary migration can, and does occur. At present, there are substantial arguments both in favor and against {\it in situ} formation of close-in ungiants, and for hot jupiters, where initial formation at large astrocentric distance seems somewhat more assured, it is not clear whether the LKCTF mechanism or quiescent  long-distance Type-II disk migration is the dominant process.

\item
{\it Why is our inner Solar System empty?} The \ik Mission and the high-precision Doppler Velocity Surveys have shown that the default mode of planet formation generates multiple super-Earth mass planets with orbital periods, $P<100\,{\rm d}$. Our own Solar System, however, contains nothing interior to Mercury's 88-day orbit. Furthermore, given the {\it presence} of Jupiter, with a mass and period combination that appear to be at least somewhat unusual, is there a possible connection between these two abnormal Solar System attributes? Work by \citep{2011Natur.475..206W} suggests that both Jupiter and Saturn experienced significant radial migration, and the so-called ``Nice Model'' \citep{ 2005Natur.435..466G, 2005Natur.435..462M, 2005Natur.435..459T} posits that substantial migration of the outer planets occurred during the solar system's formation phases. An interesting, and as yet unanswered, question is whether such a process -- generated, for example, via sweeping mean-motion resonances -- could have thwarted planet formation in our inner Solar System, or destroyed planets that were originally there. 
\end{itemize}

{\bf Acknowledgement:} We are pleased to acknowledge support from the NASA TESS Mission through subaward \# 5710003702
\bibliography{shear}

\end{document}